\documentclass[pre,twocolumn]{revtex4-1}
\usepackage{natbib}		
\usepackage{graphicx}	
\usepackage{amsmath}
\usepackage{bm}
\usepackage{xcolor}
\usepackage{textcomp,mathcomp}
\renewcommand{\vec}[1]{{\bf{}#1}}

\newcommand{\comment}[1]{}
\begin{document}

\title{Capillary interactions, aggregate formation and the rheology of particle-laden flows: a lattice Boltzmann study} 
\author{Lei Yang}
\author{Marcello Sega}
\affiliation{Helmholtz Institute Erlangen-N\"urnberg for Renewable Energy (IEK-11), Forschungszentrum J\"ulich, F\"urther Stra{\ss}e 248, 90429 N\"urnberg, Germany}
\author{Steffen Leimbach}
\author{Sebastian Kolb}
\author{J\"urgen Karl}
\affiliation{Chair of Energy Process Engineering, Friedrich-Alexander-Universit\"at Erlangen-N\"urnberg,
Fürther Stra{\ss}e 244f, 90429 N\"urnberg, Germany}
\author{Jens Harting}
\affiliation{Helmholtz Institute Erlangen-Nürnberg for Renewable Energy (IEK-11), Forschungszentrum Jülich, Fürther Stra{\ss}e 248, 90429 Nürnberg, Germany}
\affiliation{Department of Chemical and Biological Engineering and Department of Physics, Friedrich-Alexander-Universit\"at Erlangen-N\"urnberg, Fürther Stra{\ss}e 248, 90429 N\"urnberg, Germany} 

\begin{abstract}
\noindent{}
The agglomeration of particles caused by the formation of capillary bridges 
has a decisive impact on the transport properties of a variety of at a first sight very different systems such as capillary suspensions, fluidized beds in chemical reactors, or even sand castles. 
Here, we study the connection between the microstructure of the agglomerates and the rheology of fluidized suspensions using a coupled lattice Boltzmann and discrete element method approach. We address the influence of the shear rate, the secondary fluid surface tension, and the suspending liquid viscosity. The presence of capillary interactions promotes the formation of either filaments or globular clusters, leading to an increased suspension viscosity. Unexpectedly, filaments have the opposite effect on the viscosity as compared to globular clusters, decreasing the suspension viscosity at larger capillary interaction strengths. In addition, we show that the suspending fluid viscosity also has a non-trivial influence on the effective viscosity of the suspension, a fact usually not taken into account by empirical models. 
\end{abstract}
\maketitle

\section{Introduction}
\begin{figure}
\begin{center}
        \includegraphics[width=0.8\columnwidth]{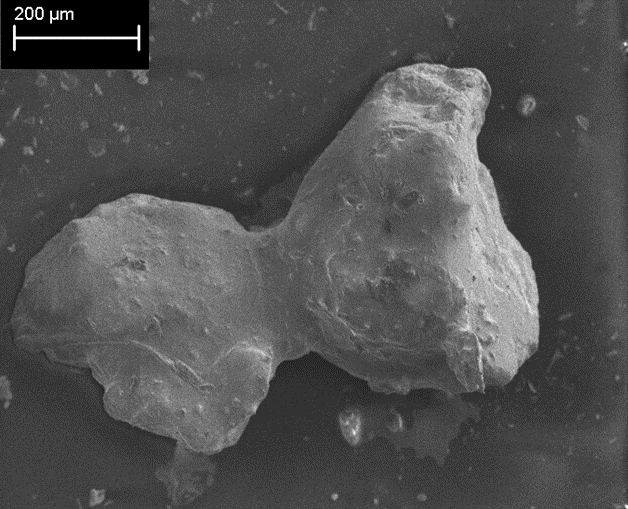}
\end{center}
\caption{SEM micrograph of agglomeration betweeen two sand particles in hot gas experiments (830\textdegree C) with sand as bed material and synthetic straw ash.}
  \label{fig:agglomeration}
\end{figure}

Particulates are often encountered in
applications in the petroleum, food, chemical, and energy industries. Their presence, thanks to the sensitivity of their collective properties to changing  external conditions, can influence to a large extent the macroscopic behavior of the medium in which they are embedded. Understanding the effects of particulates on their environment can be a complex task, especially in the presence of several components. 
A prominent example is the combustion of biomasses in fluidized beds. 
It is reported that the dominant agglomeration mechanism during the combustion of biomass is coating-induced agglomeration~\cite{visser2004influence,gatternig2015investigations}. The ashes of biogenic fuels always contain a high amount of silicon and potassium, prone to melt at operating temperatures. 
The ash melts then attach to the bed particles and form a sticky coating on their surfaces~\cite{khan2009biomass}. Favorable wettability of the ash melts coating on particle surface promotes the formation of capillary bridges~\cite{ergudenler1993agglomeration, ohman2000bed, olofsson2002bed, gatternig2015investigations}, as seen in Fig.~\ref{fig:agglomeration}. These bridges can cause the formation of a space-spanning network~\cite{koos2011capillary,koos2012tuning,herminghaus2005} and the agglomeration of particles into large clusters~\cite{ergudenler1993agglomeration, ohman2000bed, olofsson2002bed, gatternig2015investigations}. If not appropriately counteracted, the growth of agglomerates can significantly alter the suspension's fluid-dynamic properties and cause, eventually, the defluidization of the bed~\cite{cho2006defluidization, fryda2008agglomeration,scala2006combustion}. 
In industrial reactors, the occurrence of defluidization might require a plant shutdown to avoid damage. For these reasons, understanding the effect of liquid capillary bridges on the rheological properties of suspended particles is the first step to optimize the performance of fluidized bed reactors.

\comment{

}
An increase of apparent viscosity, inadequate mixing, uneven
temperature distribution and segregation are all characteristics of the defluidization~\cite{schugerl1971rheological}.
Such a behavior is very general and can be observed in a wide selection of at a first glance very different systems ranging from the above-mentioned fluidized beds to classical colloidal suspensions. 
Generally, the agglomeration of particles due to the formation of liquid bridges can give rise to a shift in the bulk rheological behavior of the suspension from nearly Newtonian, for moderate volume fractions, to non-Newtonian~\cite{van1975rheology}.  For example, Koos and Willenbacher~\cite{koos2011capillary} reported agglomeration due to capillary forces that dramatically shifted the bulk rheological behavior in a suspension of hydrophobically modified calcium carbonate in diisononyl phthalate. McCulfor and coworkers~\cite{mcculfor2011effects} also reported a largely increased suspension viscosity for glass particle suspensions in mineral oil due to particle agglomeration induced by liquid bridges. The relationship between the microstructure and the rheological behavior of such a capillary suspension system is, however, still not well understood~\cite{scala2006combustion}.

Coupled Computational Fluid Dynamics - Discrete Element Method (CFD-DEM) simulations can resolve single grains and take adhesion forces explicitly into account, representing an excellent choice to study agglomeration formation and link the microstructure with the rheology. Effective interactions that aim at a realistic description of liquid bridge-induced forces between particles allow bypassing the computational complexity of simulating explicitly the liquid bridges in CFD-DEM~\cite{mikami1998numerical,wu2018effect,zhang2017assessment,roy2017}.  Recently, we compared systematically available analytical models for the liquid bridge force with explicit liquid-bridge simulations based on a multi-component lattice Boltzmann model. In this way we identified the best-suited model for the effective description of capillary forces. In this work we make use of this model to perform coupled lattice Boltzmann and DEM simulations of the formation of agglomerates in the presence of capillary interactions. We investigate the influence of these interactions on the microstructure and rheology of the suspension to elucidate some fundamental aspects of the local rheology in fluidized beds.

The outline of the paper is as follows. In Sec.~\ref{sec:methods} we describe
the lattice Boltzmann method coupled to the DEM. In Sec.\ref{sec:results}, we investigate
the influence of several key simulation parameters (suspending liquid viscosity, surface tension, and shear rate) on the
rheological properties of the suspension. We also compare our
observations with the experimental data reported by Koos and
coworkers~\cite{koos2014restructuring}. Finally, we summarize our results. 

\section{Model description\label{sec:methods}}
The lattice Boltzmann method provides an approximate solution to the
Navier-Stokes equations in the limit of small Mach and Knudsen
numbers by computing the moments of the Boltzmann transport equation
solved on a lattice~\cite{benzi1992lattice}
\begin{equation}\label{eq:3}
   f_i (\vec{x} + \vec{c}_i \Delta t, t+\Delta t ) = f_i (\vec{x}, t ) + \Omega_i(\vec{x}, t ),
\end{equation}
where $f_i (\vec{x}, t )$ represents the  distribution function at
position  $\vec{x}$  and time $t$ of particles with velocity
$\vec{c}_i (i = 1,\ldots,N)$ commensurate with the three-dimensional
lattice (here, we use the D3Q19 velocity set with $N=19$ directions). In the following, we make use of a reduced set of units such that the lattice constant, the integration time  step $\Delta t$ and the and mass of the fluid particles are all equal to one. The Bhatnagar-Gross-Krook (BGK)
collision operator~\cite{bhatnagar1954model}
\begin{equation}\label{eq:4}
    \Omega_i = -\frac{f_i(\vec{x}, t ) - f_i^\mathrm{eq}[\rho(\vec{x},t), \vec{u}(\vec{x}, t) ] }{\tau_\mathrm{relax}},
\end{equation}
describes the relaxation of the distribution functions towards the
local Maxwell-Boltzmann equilibrium distribution, approximated as
\begin{equation}\label{eq:5}
    f_i^\mathrm{eq} = \zeta_i \rho \left[1 + \frac{\vec{c}_i\vec{u}}{c_s^2} + \frac{(\vec{c}_i\vec{u})^2}{2c_s^4} - \frac{u^2}{2c_s^2} \right].
\end{equation}
Here, $\tau_\mathrm{relax}$ is the relaxation time, $\rho(\vec{x},t)
= m \sum_{i} f_i (\vec{x},t)$ is the fluid mass density, $\vec{u} (\vec{x},t)
=  \rho^{-1} (\vec{x},t) \sum_i f_i^c (\vec{x},t) \vec{c}_i$  is
the macroscopic fluid velocity. $m$ is a reference density. $c_s = 1/\sqrt{3}$ is the speed of
sound and $\zeta_i$ are appropriate coefficients depending on
the velocity space discretization. The resulting kinematic viscosity
of the fluid is $\nu = c_s^2 (\tau_\mathrm{relax} - 1/2)$.

The solid particles are discretized on the fluid lattice and their
interaction with the fluid  is realized by means of a modified
bounce-back boundary condition acting on the surface of the
particle~\cite{ladd1994numerical,aidun1998direct}
\begin{equation}\label{eq:6}
     f_i^c (\vec{x} + \vec{c}^*_i, t+1 ) = f_i^c (\vec{x}+\vec{c}_i, t ) + \Omega_i^c (\vec{x}, t )+ \mathcal{C},
\end{equation}
where $\vec{c}^*$ is the reflected direction and
$\mathcal{C}=2c_s^{-2}\zeta_i \rho^c(\vec{x}+\vec{c}_i,t)\vec{v}_b
\cdot \vec{c}_i $ is a first-order velocity correction with $\vec{v}_b$
the local particle surface velocity. To compensate for the change of momentum
of the fluid caused by the bounce-back Eq.~\ref{eq:6}, the
reaction force
\begin{equation}
     \vec{F}(t) = [2 f_i^c (\vec{x}+\vec{c}_i, t ) + \mathcal{C}] \vec{c}_i
\end{equation}
needs to act on the particle.

The motion of the particles follows the rigid body equations of motion
\begin{align}
    \vec{F}_p =& m_p \frac{d \vec{u}_p}{dt},\\
    \vec{T}_p   =& J \frac{d \bm{\omega}_p}{dt},
\end{align}
where $\vec{F}_p$ and $\vec{T}_p$ are the total force and torque
acting on particle $p$ of mass $m_p$, velocity $\vec{u}_p$, moment of inertia
$J$ and angular velocity $\bm{\omega}_p$.

When the surfaces of two particles of radius $R_p$ are getting closer than the
cut-off distance $r_c = 2/3$, we apply a lubrication correction in the
form of a pair interaction as proposed by Ladd and
Verberg~\cite{ladd2001lattice}
\begin{equation}
  \vec{F}_{ij}^\mathrm{lub} =  \frac{3 \pi \mu R_p^2}{2} \hat{r}_{ij} \left( \hat{r}_{ij} \cdot \vec{u}_{ij} \right) \left( \frac{1}{r_{ij} - 2R_p} - \frac{1}{r_c}    \right),
\end{equation}
where $\mu=\rho \nu$ is the dynamic viscosity of the fluid, and
$r_{ij}$ is the distance between the centers of particles $i$ and $j$, $\hat{r}_{ij}$ is the corresponding unit vector, and $\vec{u}_{ij} = \vec{u}_{i} - \vec{u}_{j}$ is the relative velocity of the two particles.
Besides, we use the Hertz
hard core potential $U_H$ between particles at close
contact~\cite{hertz1882ueber}, which, for two identical particles, is given by
\begin{equation}
    U_H= K_H S^{5/2} \textrm{  for  } r_{ij} < 2R_p,
\end{equation}
where here we use the force constant $K_H=100$, and the  separation distance $S$ between the spheres $i$ and $j$ is
\begin{equation}
    S = 2 R_p -  r_{ij}.
\end{equation}

The capillary bridge forces are modelled in an effective
way, using the pair interaction of  Willett and
coworkers~\cite{willett2000capillary}
\begin{equation}
\vec{F}_{ij}^\mathrm{cap} =   2\pi\sigma R_p  \exp\left[f_1(\theta,V) - f_2(\theta,V)\exp(g)\right]  \hat{r}_{ij}
\end{equation}
\begin{equation}
g = f_3(\theta,V) \ln\frac{S}{2\sqrt{V/R_p}} + f_4(\theta,V) \ln^2\frac{S}{2\sqrt{V/R_p}},
\end{equation}
where the volume of the capillary bridge 
which we
showed~\cite{yang2020capillary} to  be an excellent approximation
of the capillary force acting between two spherical particles in
the presence of a fluid bridge described explicitly with a secondary fluid phase.
The definitions of the functions $f_1, f_2, f_3$ and $f_4$ are reported in the Appendix.
Here, $\sigma$ is the liquid surface tension, $\theta$ is the contact angle, and $V$ is the volume of secondary fluid present on the particle.
The capillary force is applied only if the particle gap is smaller than the rupture distance $S_\mathrm{rup}$, which, in turn, depends on $V$ in the following approximate way~\cite{willett2000capillary}
\begin{equation}
    S_\mathrm{rup} \simeq R_p \left(1+ \frac{\theta}{2} \right) \left( \frac{V^{1/3}}{R_p} + \frac{1}{10}\frac{V^{2/3}}{R_p^2} \right).
\end{equation}

Similarly to Mikami
and coworkers~\cite{mikami1998numerical}, we assume that the secondary
fluid is, effectively, uniformly distributed among all the particles.
In addition, we neglect liquid transport between particles and any
change of liquid bridge volume or shape by the flow. In all
simulations, we use a constant liquid volume fraction  $V/V_p =  5\times{}10^{-3}$, namely, the ratio of the 
volumes $V$ of the secondary fluid involved in a bridge and that of a particle,  $V_p$. In addition we choose the value of the particle volume fraction 
($\phi=0.11$) and of the contact angle ($\theta=123$\textdegree) to match the 
experimental conditions in the work by Koos and 
coworkers~\cite{koos2014restructuring}.
We model different capillary bridge strength by
changing the value of the surface tension in the range from $5\times{}10^{-5}$ to $5\times{}10^{-4}$. Besides, we perform simulations at different values of the suspending medium viscosity in the range from 1/15 to 1/6.

We study the rheology of the suspension by imposing a shear flow
using Lees-Edwards boundary conditions~\cite{lees1972computer,wagner2002lees,HVC04}
along the $x$-axis, generating a spatially homogeneous linear
shear flow along the $z$-axis. The Lees-Edwards boundary conditions for particles are implemented to allow the consistent treatment of solid particles
crossing the boundary, where the shear stress $\tau$ is computed. To prevent
the unavoidable center of mass drift due to roundoff errors, the
center of mass velocity is reset to zero at prescribed time intervals,
without influencing the evolution of the
simulation~\cite{lallemand2000theory,wagner2002lees}. The effective
shear rate $\dot\gamma_\mathrm{eff}$ is obtained by a linear fit of the flow
velocity profile. Similar to the work of Huang and
coworkers~\cite{huang2012rotation} and
Janoschek~\cite{janoschek2013mesoscopic}, the total shear stress
$\tau$ is measured by time-averaging the local shear stress acting
on the fluid nodes next to the plane where Lees-Edwards boundary conditions are applied. The effective viscosity
is then calculated as
\begin{equation}
    \mu_\mathrm{eff} = \frac{\tau}{\dot\gamma_\mathrm{eff}}
\end{equation}
and the relative suspension viscosity is calculated as $\mu_r = \mu_\mathrm{eff}/\mu$.
Initial configurations are generated by placing randomly 2041
identical particles of radius $R_p=3$ in a cubic simulation box of
edge length 128, corresponding to volume packing fractions
11\%. 

\section{Results and discussion\label{sec:results}}
\begin{figure*}[t]
\raisebox{0.15\columnwidth}{\parbox{0.2\columnwidth}{$\dot\gamma_\mathrm{eff}=1.4\times10^{-6}$}}\includegraphics[trim=160 50 60 80, clip,width=0.35\columnwidth]{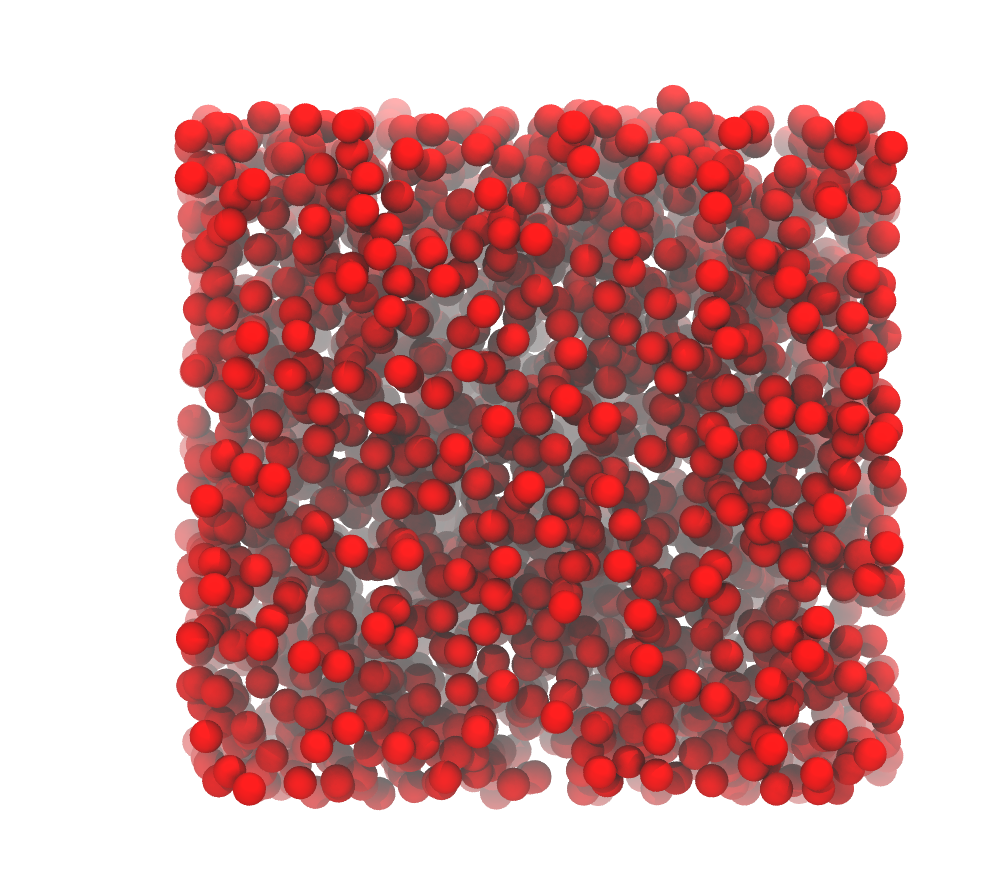}
\includegraphics[trim=160 50 60 80, clip,width=0.35\columnwidth]{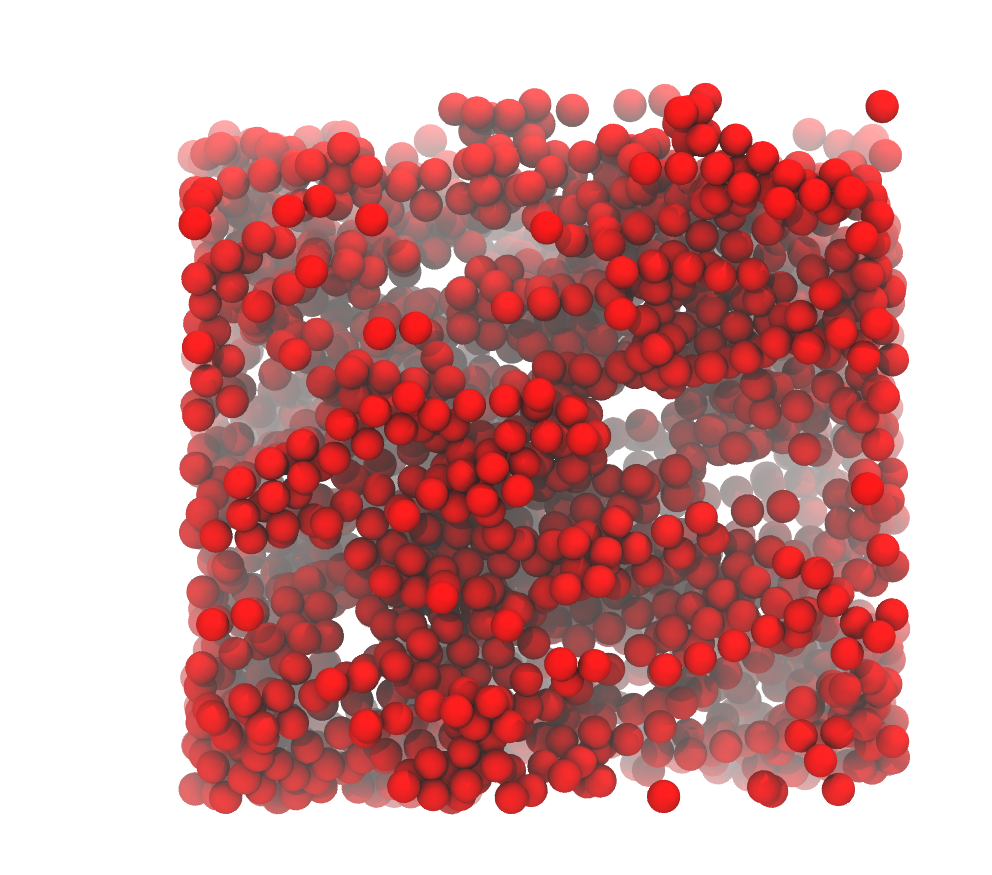}
\includegraphics[trim=160 50 60 80, clip,width=0.35\columnwidth]{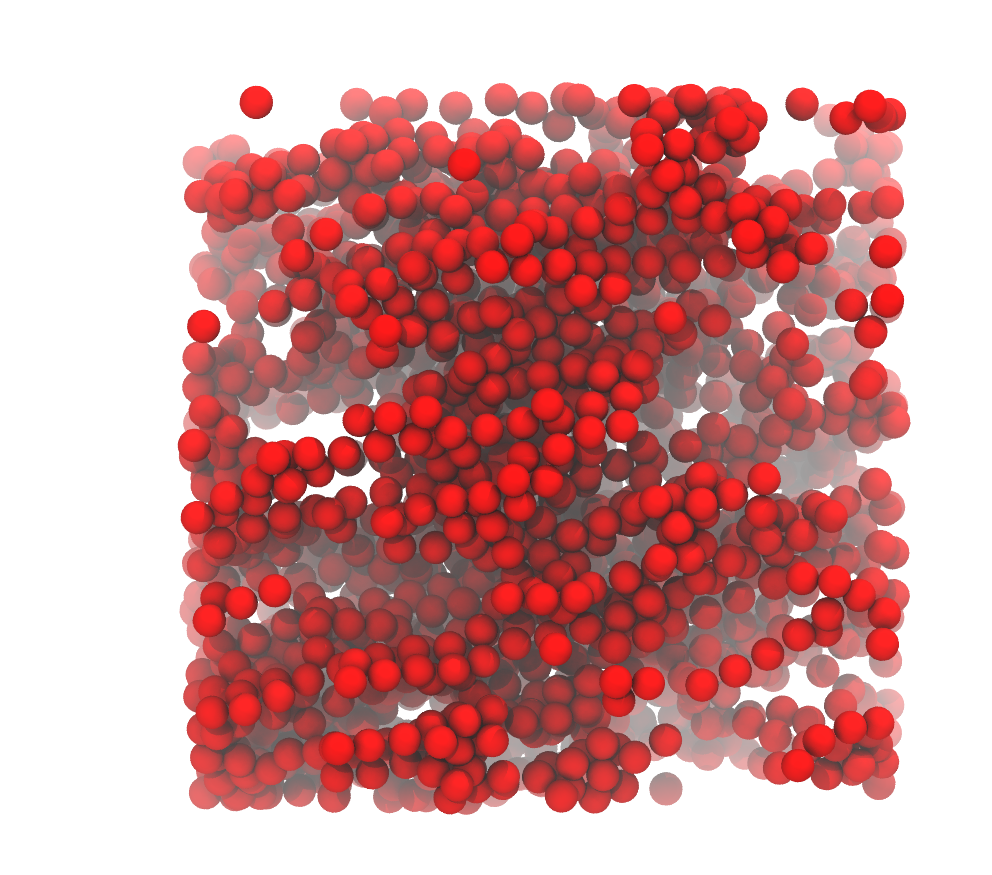}
\includegraphics[trim=160 50 60 80, clip,width=0.35\columnwidth]{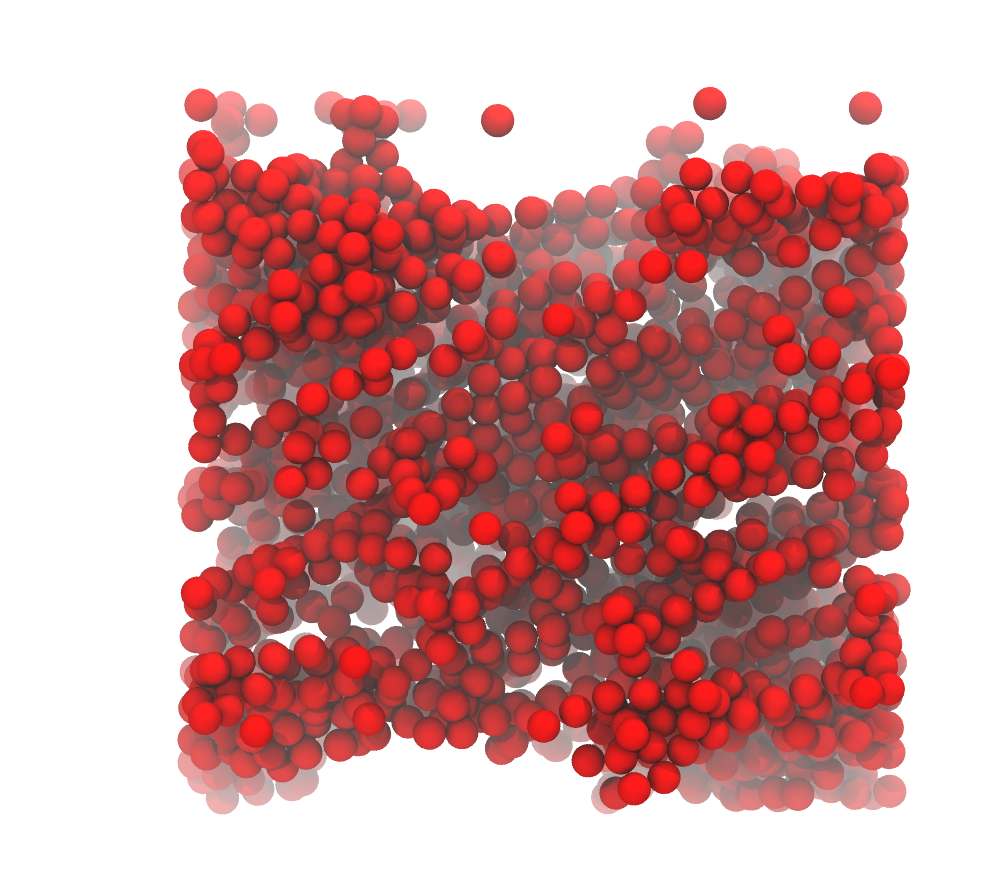}\\
\parbox{0.2\columnwidth}{\phantom{M}}
\parbox{0.35\columnwidth}{$\sigma=0$}
\parbox{0.35\columnwidth}{$\sigma=5\times10^{-5}$}
\parbox{0.35\columnwidth}{$\sigma=1.5\times10^{-4}$}
\parbox{0.35\columnwidth}{$\sigma=5\times10^{-4}$}
\\
\raisebox{0.15\columnwidth}{\parbox{0.2\columnwidth}{$\dot\gamma_\mathrm{eff}=4.5\times10^{-5}$}}\includegraphics[trim=130 80 130 80, clip,width=0.35\columnwidth]{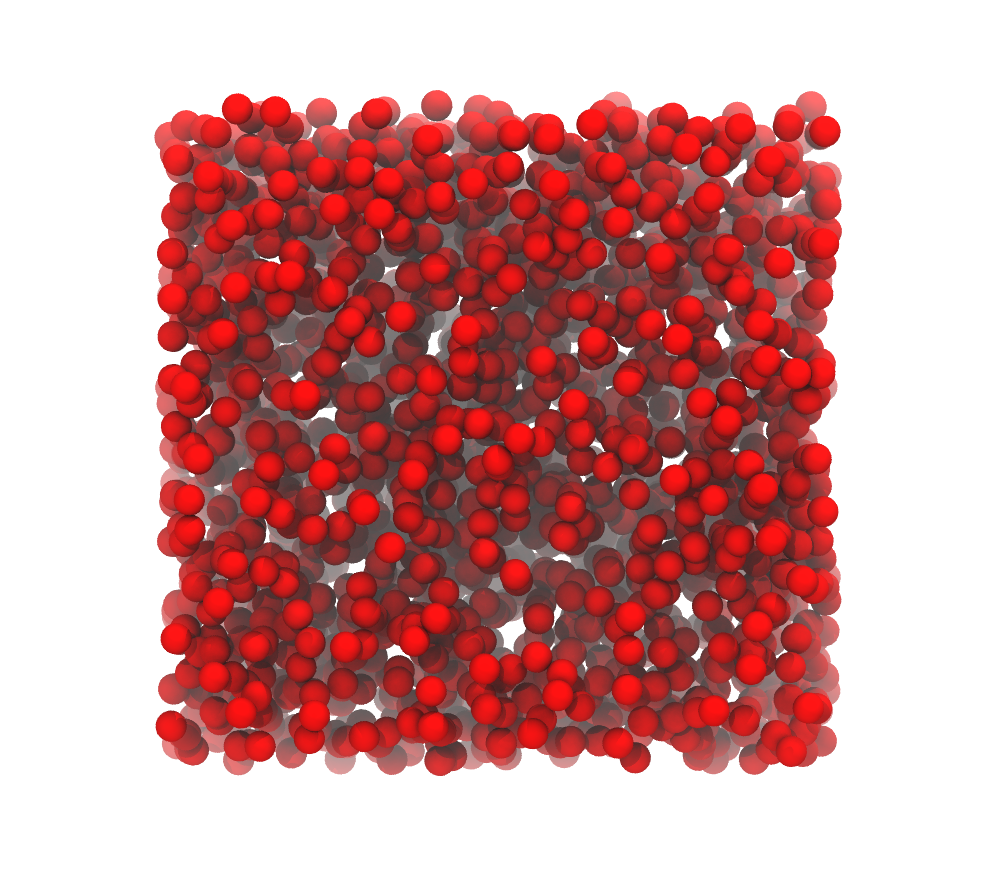}
\includegraphics[trim=130 80 130 80, clip,width=0.35\columnwidth]{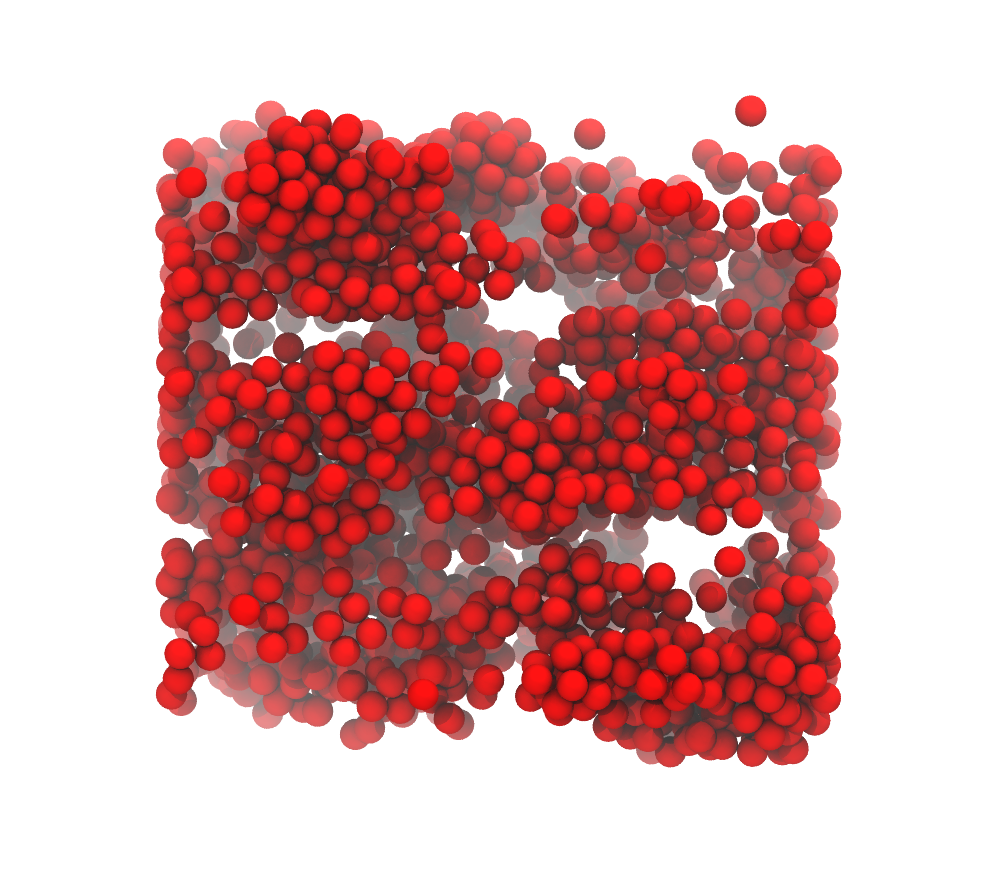}
\includegraphics[trim=130 80 130 80, clip,width=0.35\columnwidth]{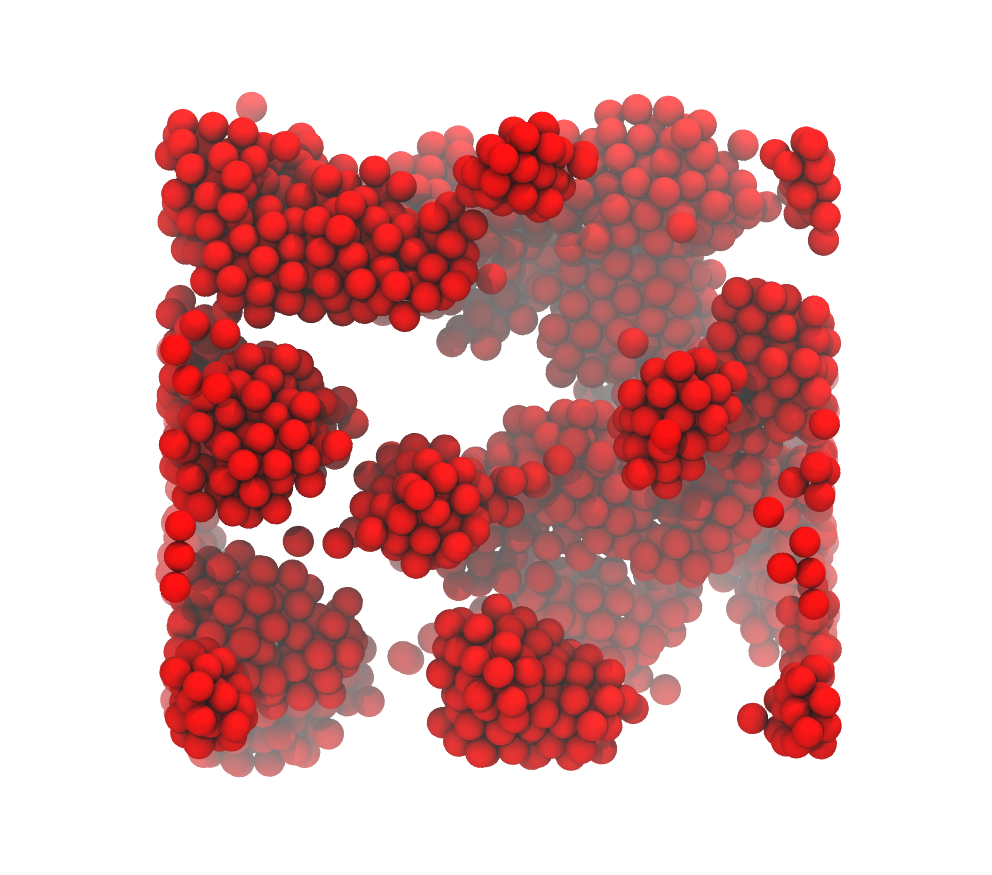}
\includegraphics[trim=120 80 130 80, clip,width=0.35\columnwidth]{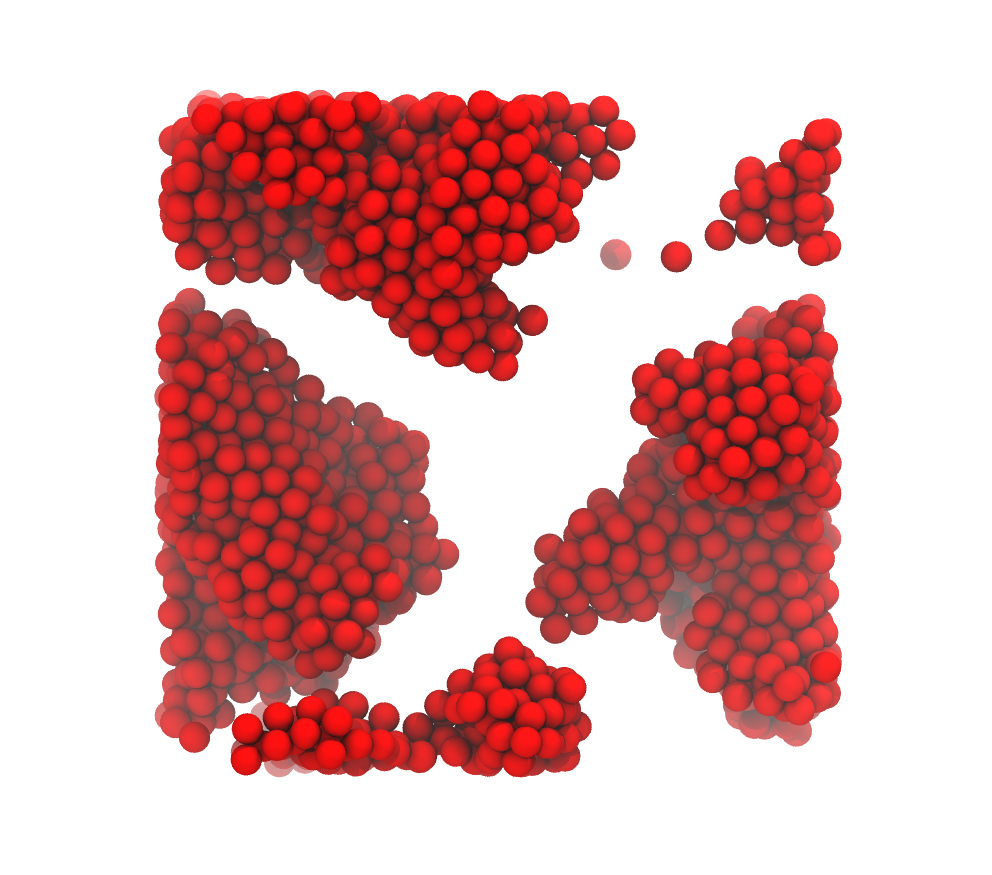}\\
\raisebox{0.15\columnwidth}{\parbox{0.2\columnwidth}{$\sigma=1.5\times10^{-4}$}}\includegraphics[trim=160 50 60 80, clip,width=0.35\columnwidth]{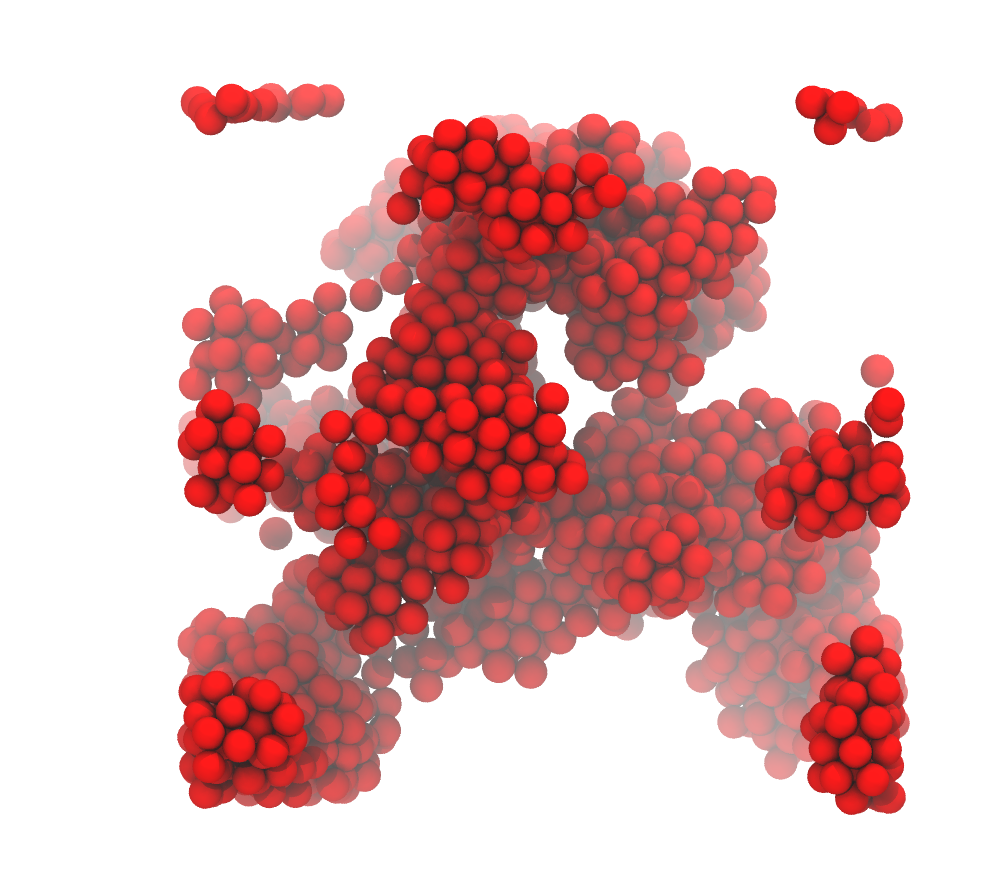}
\includegraphics[trim=160 50 60 80, clip,width=0.35\columnwidth]{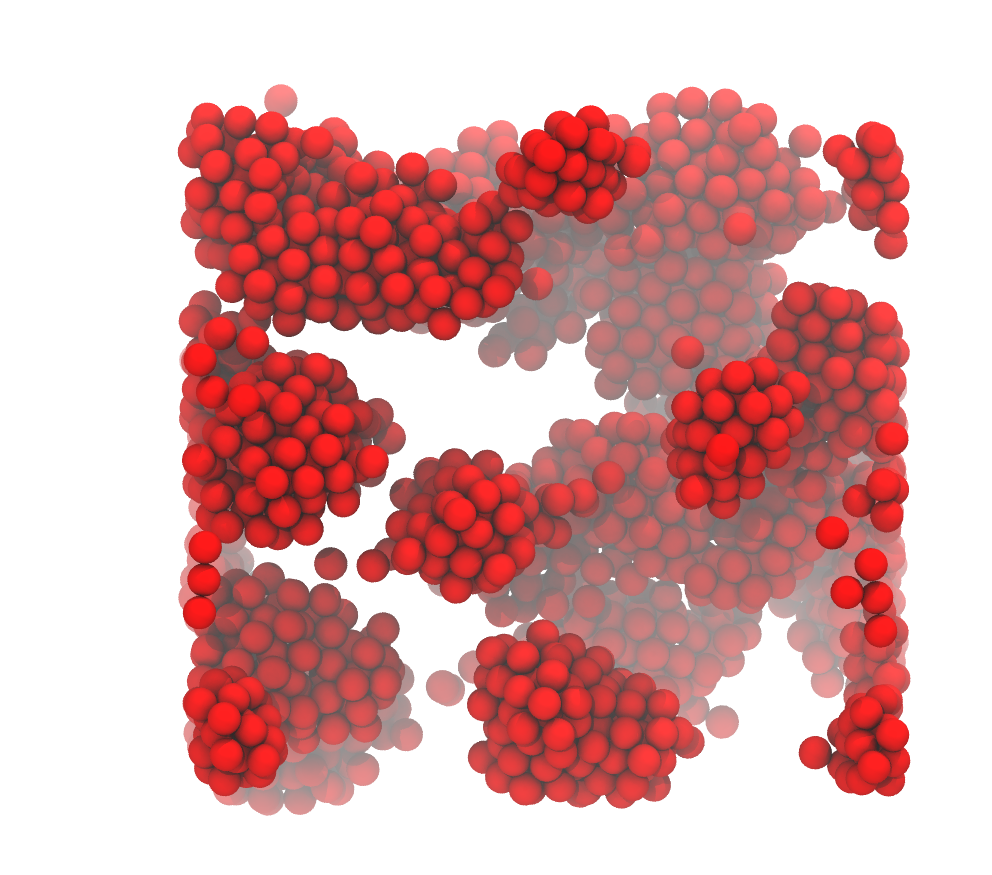}
\includegraphics[trim=160 50 60 80, clip,width=0.35\columnwidth]{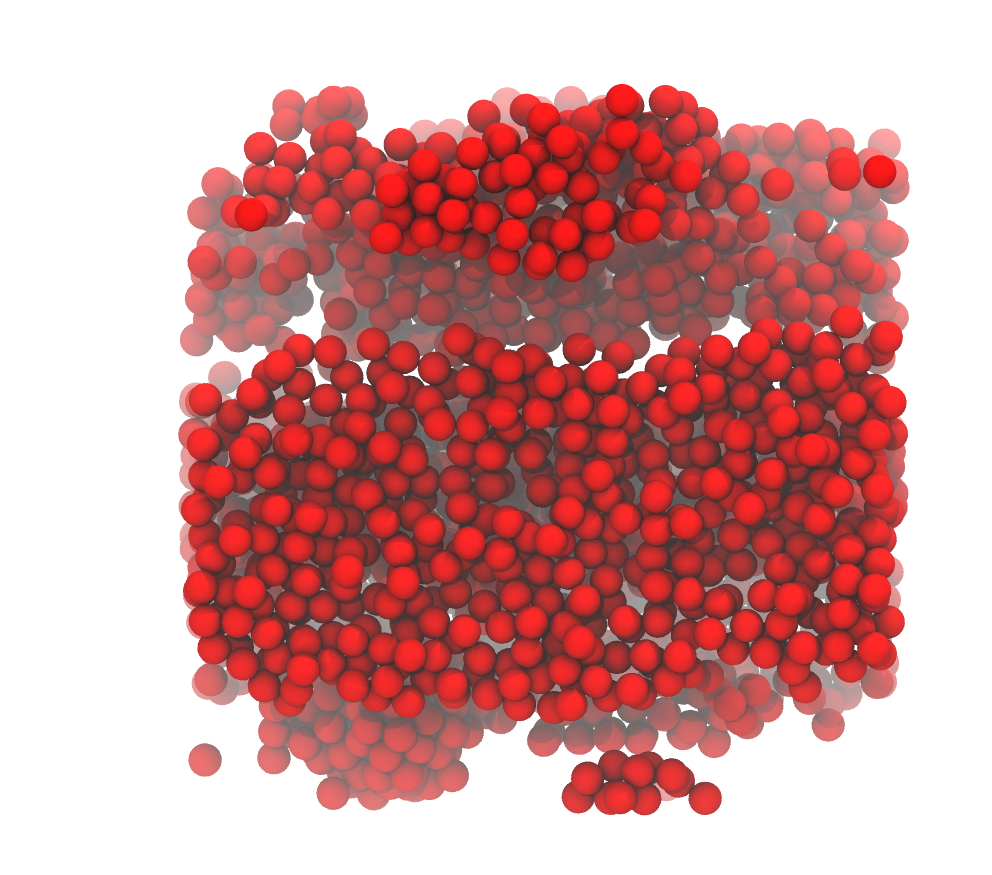}
\includegraphics[trim=160 50 60 80, clip,width=0.35\columnwidth]{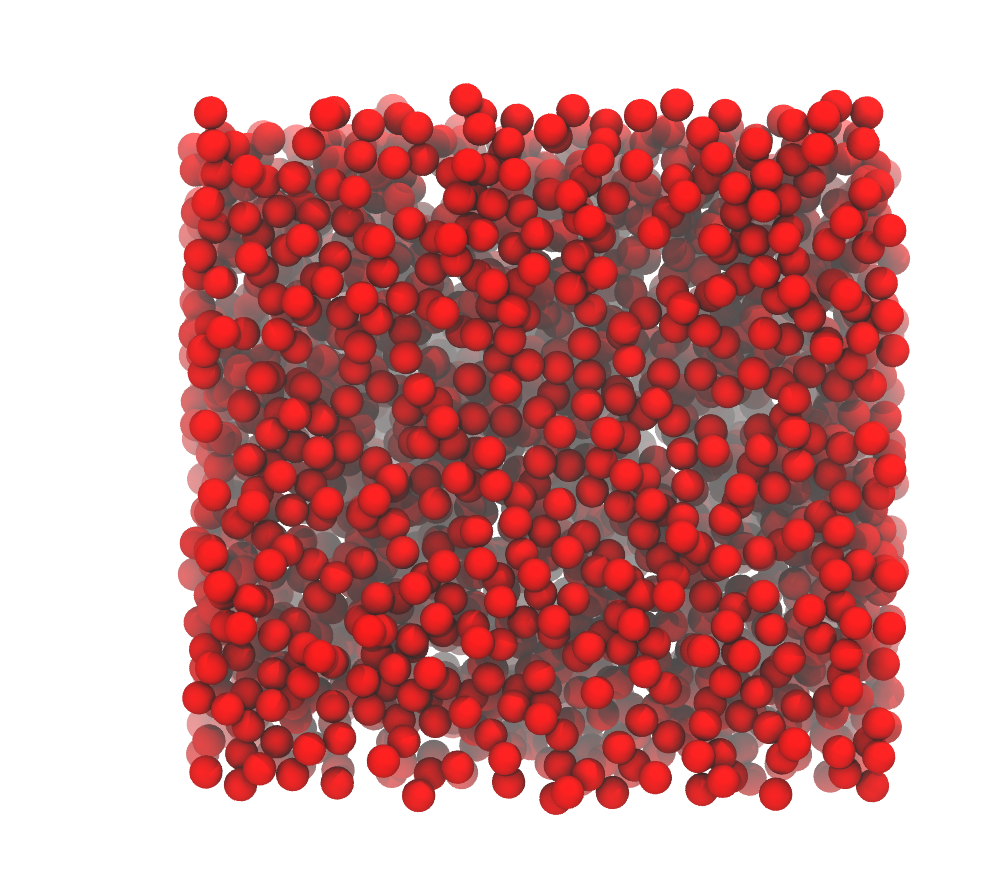}\\
\parbox{0.2\columnwidth}{\phantom{M}}
\parbox{0.35\columnwidth}{$\dot\gamma_\mathrm{eff}=1.4\times10^{-5}$}
\parbox{0.35\columnwidth}{$\dot\gamma_\mathrm{eff}=4.5\times10^{-5}$}
\parbox{0.35\columnwidth}{$\dot\gamma_\mathrm{eff}=1.4\times10^{-4}$}
\parbox{0.35\columnwidth}{$\dot\gamma_\mathrm{eff}=4.2\times10^{-4}$}
    \caption{Simulation snapshots of particle clustering in the the stationary shear flow. Top row:  fixed $\dot\gamma_\mathrm{eff}=1.4\times10^{-6}$ and increasing capillary force strength from left to right: no capillary force, $\sigma=5\times10^{-5}$,$1.5\times10^{-4}$, and $5\times10^{-4}$. 
    Center row: fixed $\dot\gamma_\mathrm{eff}=4.5\times10^{-5}$ and capillary force strength as above. 
    Bottom row: fixed $\sigma=1.5\times10^{-4}$ and  increasing effective shear rate from left to right: $\dot\gamma_\mathrm{eff}=1.4\times10^{-5},4.5\times10^{-5},1.4\times10^{-4},$ and $4.2\times10^{-4}$. 
\label{fig:snapshots}}
\end{figure*}

In the presence of capillary interactions, the most salient characteristic of the suspension is the formation of clusters, which is strongly modulated by both the applied shear and the capillary interaction strength (the surface tension value), but also in part by the suspending medium viscosity itself. The strong dependence of the aggregates as a function of the capillary interaction strength can be appreciated already by visual inspection of the simulation snapshots, as presented in Fig.~\ref{fig:snapshots}.

In the first two rows of Fig.~\ref{fig:snapshots} we present snapshots taken at constant effective shear rate (top row: $1.4\times10^{-6}$, middle row: $4.5\times10^{-5}$) and at four different capillary interaction strengths, increasing from zero to $5\times10^{-4}$ from left to right. In the bottom row, instead, we present snapshots taken at fixed capillary interaction strength ($\sigma=1.5\times10^{-4})$ and different effective shear rates, increasing from left to right.   
Visual inspection suggests the presence of three different types of structural patterns, namely (a) homogeneous distribution, (b) percolating filaments and (c) globular clusters. The particles happen to be homogeneously distributed either in absence of capillary interactions (top two rows, leftmost column of Fig.~\ref{fig:snapshots}) or at high shear rates (bottom row, rightmost column of Fig.~\ref{fig:snapshots}). The percolating filaments are clearly visible in the topmost row for all values of the capillary interaction strengths simulated, whereas the particles appear to be increasingly inhomogeneously distributed when the interaction strength increases (from left to right). A closer visual inspection shows that the filaments are also connected along the direction of the $y$-axis.  Globular clusters, finally, appear at moderately high effective shear rates. These clusters tend to be larger and more well separated with increasing capillary interaction strength (center row of Fig.~\ref{fig:snapshots}, from left to right)  and with decreasing effective shear rate (bottom row of Fig.~\ref{fig:snapshots}, from right to left).

\begin{figure}[t]
\begin{center}
    \includegraphics[width=0.98\columnwidth]{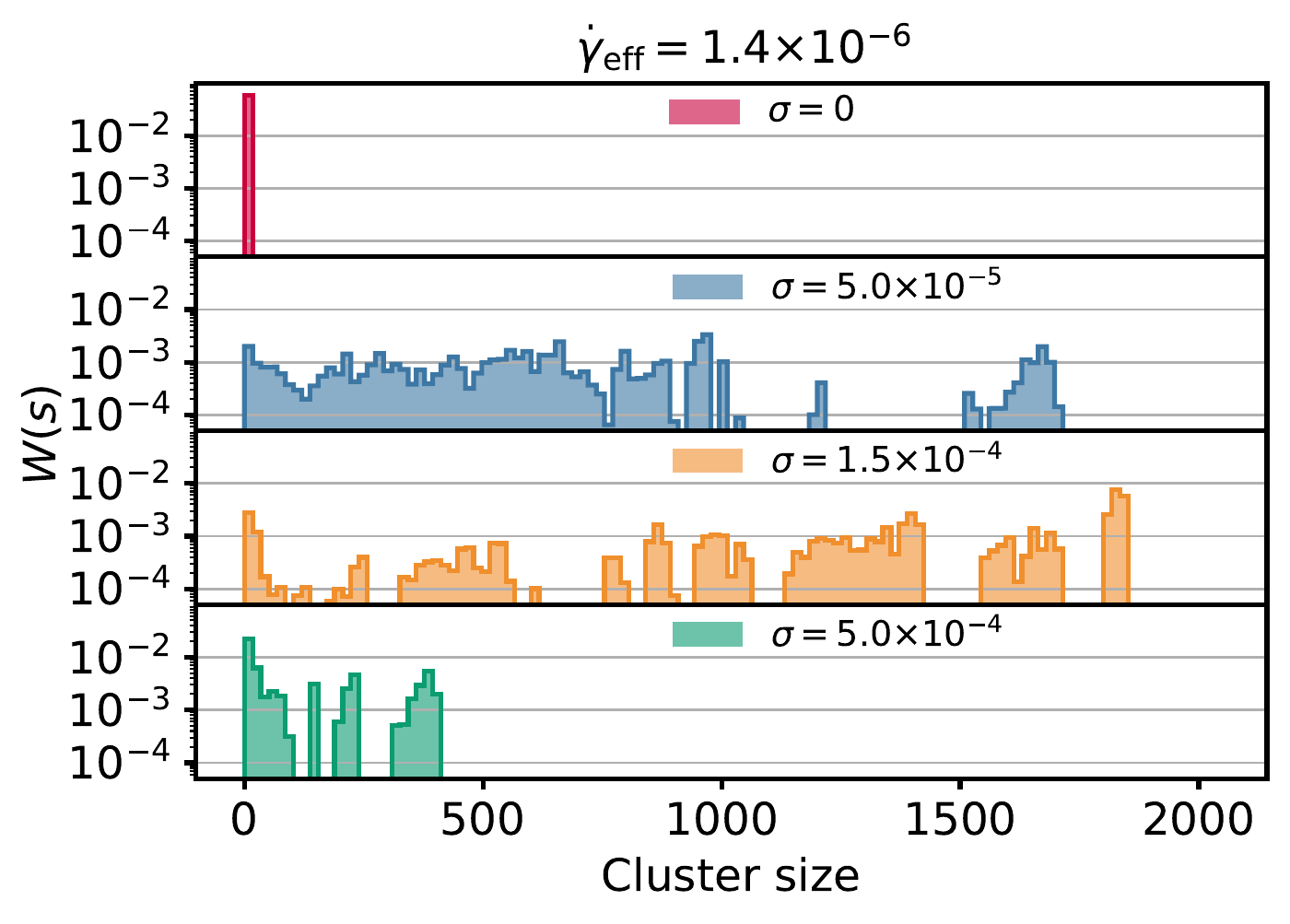}
 \end{center}
    \caption{Fraction of particles in clusters of given size  computed in the stationary state. All simulations performed      $\dot\gamma_\mathrm{eff}=1.4\times10^{-6}$ and $\mu=1/6$.\label{fig:pdf_shearrate1}}
\end{figure}

\begin{figure}[t]
\begin{center}
    \includegraphics[width=0.98\columnwidth]{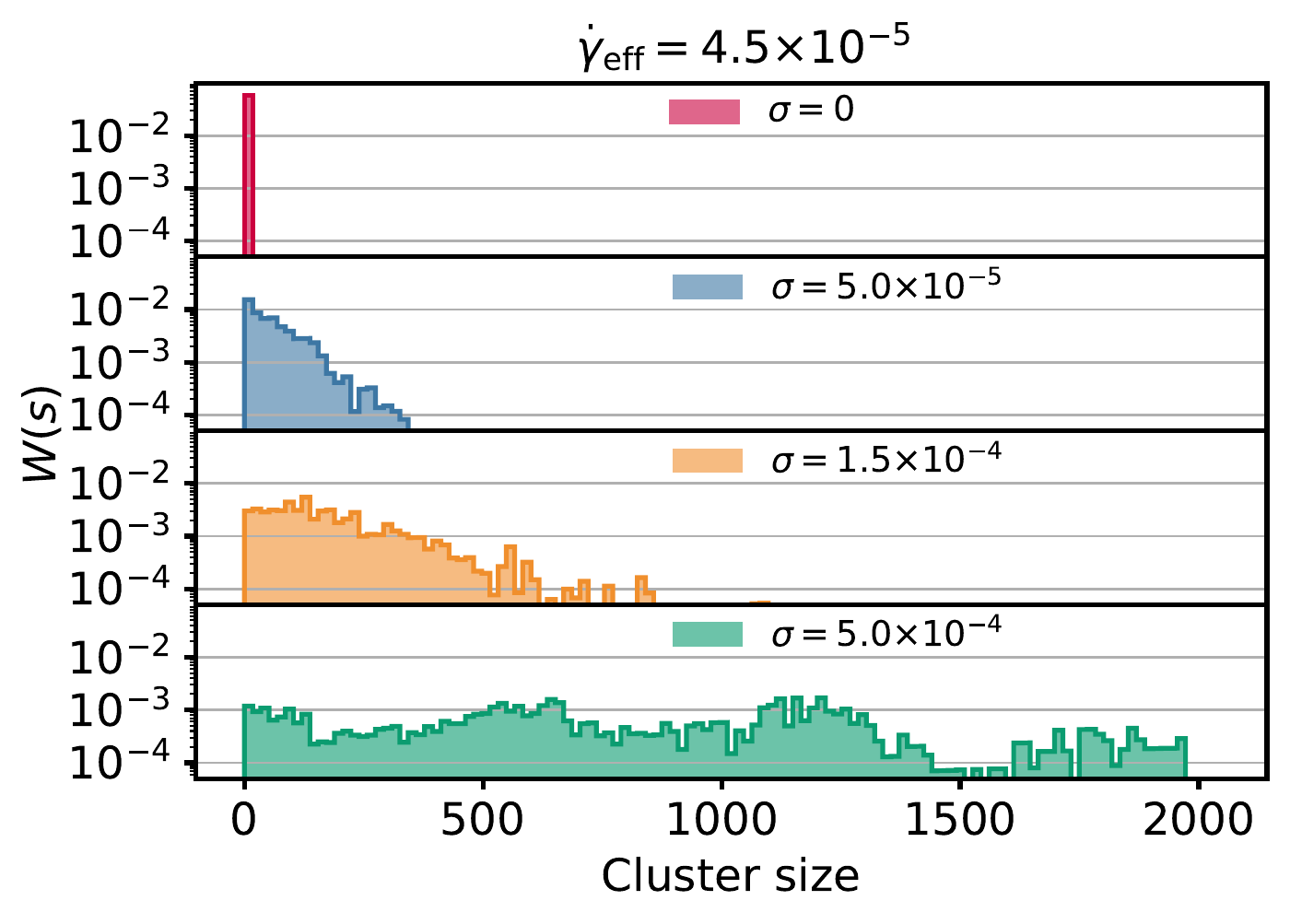}
 \end{center}
    \caption{Fraction of particles in clusters of given size  computed in the stationary state. All simulations performed at      $\dot\gamma_\mathrm{eff}=4.5\times10^{-5}$ and $\mu=1/6$.\label{fig:pdf_shearrate2}}
\end{figure}

\begin{figure}[t]
\begin{center}
    \includegraphics[width=0.98\columnwidth]{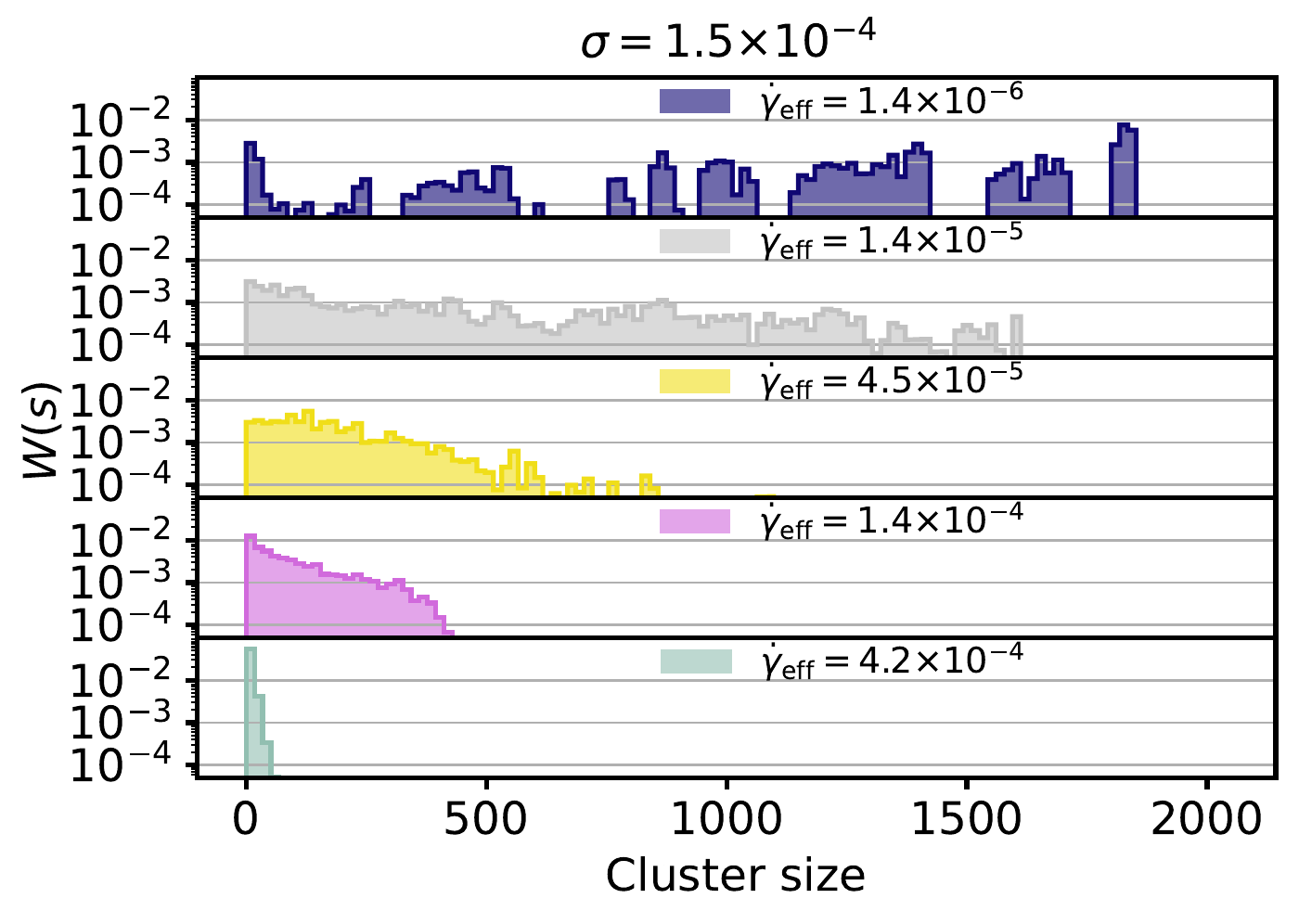}
 \end{center}
    \caption{Fraction of particles in clusters of given size  computed in the stationary state. All simulations performed at     $\sigma=1.5\times10^{-4}$ and $\mu=1/6$.\label{fig:pdf_shearrate3}}
\end{figure}

\begin{figure}[t]
\begin{center}
        \includegraphics[width=\columnwidth]{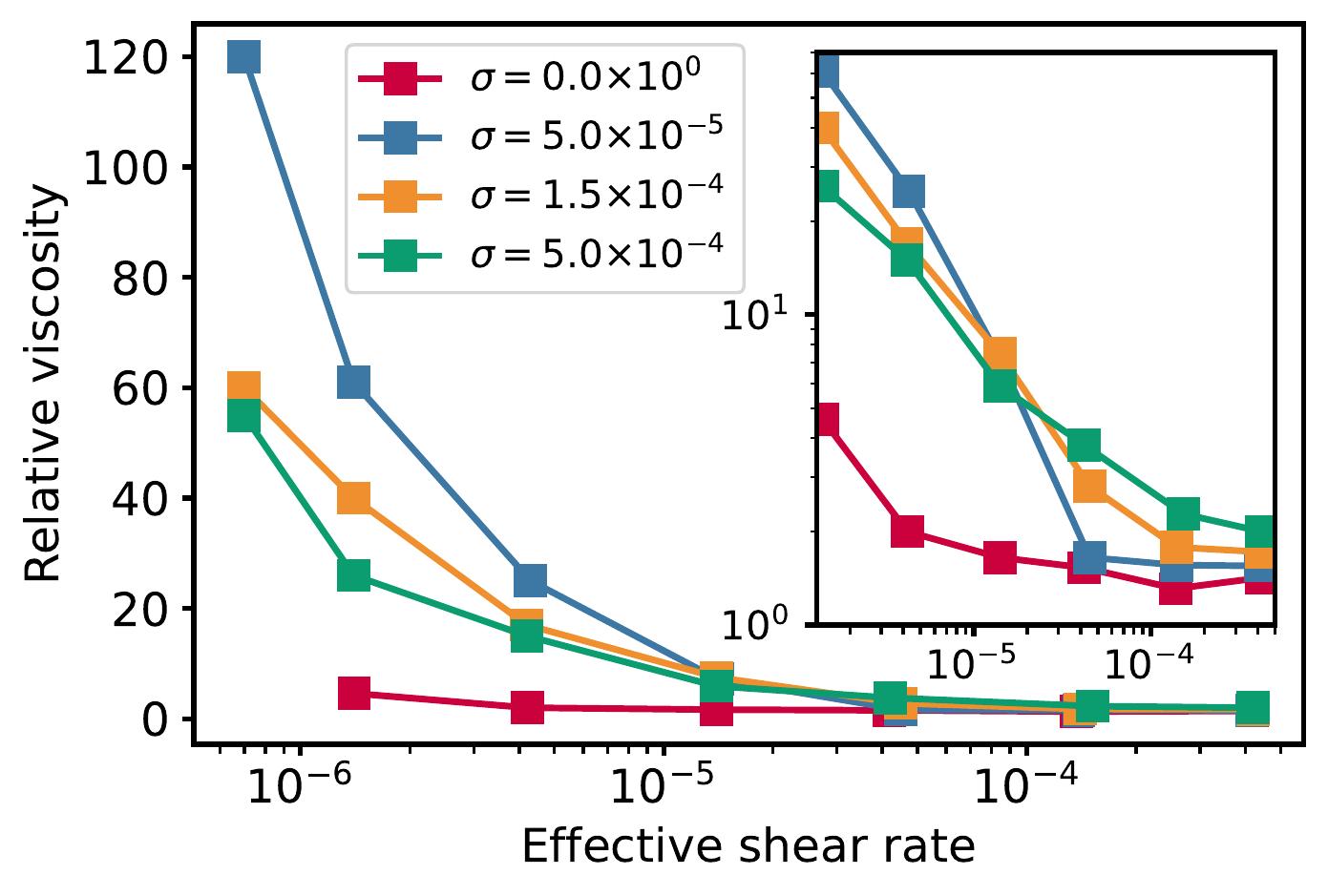}
\end{center}
\caption{Relative suspension viscosity as a function of the effective shear rate $\dot\gamma_\mathrm{eff}$. All simulations are performed at $\mu=1/6$. Inset: logarithmic scale emphasizing the differences at high shear rates. 
  \label{fig:nocap_cap_shear}}
\end{figure}

We will discuss later the impact of these structures on the rheological properties of the suspension. First, we provide a more quantitative picture of the agglomeration properties of the suspension by studying the probability of forming clusters of a given size. We begin by defining two particles as belonging to the same cluster of particles if their centers are within a given cutoff distance $\delta=6.15$, which is slightly more than the contact distance $2 R_p = 6$. The size $s$ of a cluster is simply defined as the number of particles belonging to the same cluster. Of course, $\sum_\mathrm{clusters} s_i = N_p$, where $N_p$ is the total number of particles.  Using this definition, isolated particles can be defined as clusters of size one.  We are interested in the statistical properties of the cluster sizes. However, their probability distribution function $P(s)$ is naturally skewed towards low values of $s$. It is usually more informative to look at the fraction of particles belonging to clusters of a given size, namely,
\begin{equation}
    W(s) = s P(s)/N_p.
\end{equation} 
According to the previous definition, if $P(s)$ is normalized to one, so will be $W(s)$. 
We use the Pytim software package~\cite{sega2018pytim} to determine the clusters and the corresponding distribution $W(s)$, which are presented in Figs.~\ref{fig:pdf_shearrate1}--\ref{fig:pdf_shearrate3} for different shear rates and capillary interaction strength, corresponding to the systems shown in Fig.~\ref{fig:snapshots}. The distributions show that both the capillary interaction strength and the shear rate have a profound influence on the distribution of cluster sizes. 
At low shear rates (Fig.~\ref{fig:pdf_shearrate1}), the introduction of a small capillary force ($\sigma=5\times10^{-5}$) generates promptly a large number of clusters, whereas a further increase of the capillary strength corresponds to a shift in the cluster size population towards smaller values. The trend is opposite at larger shear rates (Fig.~\ref{fig:pdf_shearrate2}), where the population of large clusters steadily increases by raising the capillary strength from $\sigma=5\times10^{-5}$ to $5\times10^{-4}$.
The distribution of particles in clusters changes roughly in a monotonous way also when keeping the interaction strength constant and increasing the shear rate (Fig.~\ref{fig:pdf_shearrate3}). At the highest effective shear rate, only clusters as large as about 60 units are found, but they involve much fewer particles than the smallest clusters (5 or less particles), which is the dominant part of the population at $\dot\gamma_\mathrm{eff}=4.2\times10^{-4}$.

The formation of various structures and the dependence of their size on the secondary fluid surface tension $\sigma$ and effective shear rate has a major impact on the suspension's relative viscosity, which we report in Fig.~\ref{fig:nocap_cap_shear}. The obvious general trend is that of a prominent shear thinning. Also, the viscosity in presence of capillary interactions never becomes smaller than its corresponding value at the same shear rate but in absence of interactions.
The shear thinning behavior presented in Fig.~\ref{fig:nocap_cap_shear} has been already reported in the literature~\cite{mcculfor2011effects,koos2012tuning,hoffmann2014using,koos2014restructuring} and has the intuitive explanation that at a low shear rate a space-filling network is formed. Here, the motion of the particles is strongly correlated, with the consequence of a significantly increased viscosity. We can show that the strong increase of viscosity at low shear rates is associated to the formation of percolating filaments. By increasing the shear rate, these filaments break, and particles regroup in disconnected or loosely connected globular clusters, loosing part of the ability to form a strongly bound, percolating network, thus decreasing the viscosity of the suspension. This is reflected in the distribution $W(s)$ reported in Fig.~\ref{fig:pdf_shearrate3}. There, the distribution that yields a high relative viscosity (at $\dot\gamma_\mathrm{eff}=1.4\times10^{-6}$) shows clearly that more particles are involved in very large clusters than at the next three higher shear rates. The progressive dissolution of these clusters at increasing shear rates is associated to the decrease in viscosity. It is important to note that even at very high shear rates, when the distribution of particles appears homogeneous in the snapshots, the presence of a small secondary volume fraction ($V/V_p = 5\times{}10^{-3}$) still yields a viscosity that is higher than in the non-interacting case, and the corresponding distribution (Fig.~\ref{fig:pdf_shearrate3}, $\dot\gamma_\mathrm{eff}=4.2\times10^{-4}$) is in fact still considerably broader than in absence of capillary interactions.
The effect of capillary bridges, therefore, can endure high shear rates, even when filaments and larger clusters disappear.

At constant shear rates, an increment in the capillary interaction strength can either decrease or increase the relative viscosity, depending whether  $\dot\gamma_\mathrm{eff}$ is smaller or larger than $1.4\times10^{-5}$.
This behavior is again consistent with the decrease (increase) in the presence of large clusters at low (high) shear rates. Qualitatively, these two opposite trends seem to be connected with the different morphology of the aggregates (filaments at low shear rates, globular clusters at high shear rates).

\begin{figure}[t]
\begin{center}
        \includegraphics[width=\columnwidth]{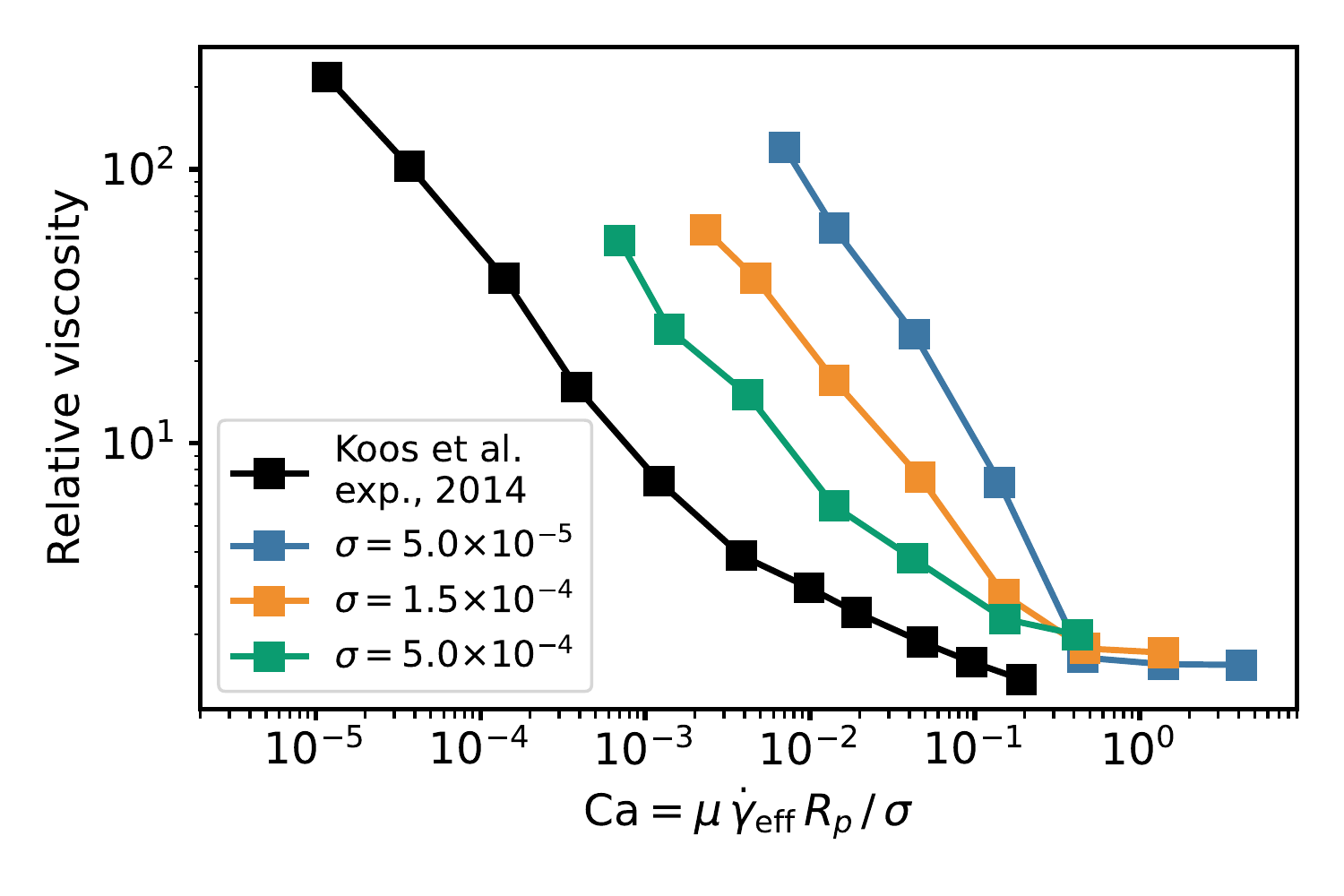}
\end{center}
\caption{Viscosity curves as a function of the capillary number. The simulation details are the same as those in Fig.~\ref{fig:nocap_cap_shear}. The black line reports the 
experimental data from Koos and coworkers~\cite{koos2014restructuring}.}
  \label{fig:vis_shearrate}
\end{figure}

To put all this in perspective and to be able to compare with experiments, we report the relative viscosity data in Fig.~\ref{fig:vis_shearrate} as a function of the adimensional capillary number $Ca=\mu \dot\gamma_\mathrm{eff}R_p/\sigma$. The results are compared with experimental measurements of Koos and
coworkers~\cite{koos2014restructuring}, which were performed using hydrophobically modified calcium carbonate
suspended in silicone oil ($\phi=0.11$, contact angle $\theta=123$\textdegree) with the addition of water ($0.1 \%\,$wt.) as secondary fluid. The volume fraction and contact angle used in our simulations are the same as in the experimental setup. When expressed in terms of the capillary number, the viscosity curves show a systematic shift to lower capillary numbers when the surface tension increases, approaching the experimental viscosity curve. 
When trying to simulate higher values of the surface tension, we always obtain a single big cluster, a typical finite size effect. In this way, we lose the ability to describe in a statistically significant way the suspension, and to perform a meaningful measurement of the effective viscosity. This issue could be avoided, in principle, by using larger simulation boxes, but this would be beyond our current capabilities. 
Additionally, we are not able to achieve even lower shear rates for systems with strong capillary forces preventing us from investigating the possible occurrence of a yield stress as observed experimentally~\cite{koos2011capillary,koos2012tuning}.
We consider the trend that approaches the experimental result as quite satisfying, taking into account the rather severe  approximations involved in the 
model. In particular, we assume that 
the  secondary fluid is distributed among  all 
particles uniformly,  which is not the case in the  
experiment.  Furthermore, we model the suspension as 
a mono-disperse  distribution of spherical  
particles, while in the experiment the particles are 
neither  spherical, nor mono-disperse. Therefore, by 
choosing the same volume fraction we match only the 
first moment of the size distribution.

\begin{figure}[t]
\begin{center}      
        \includegraphics[width=0.95\columnwidth]{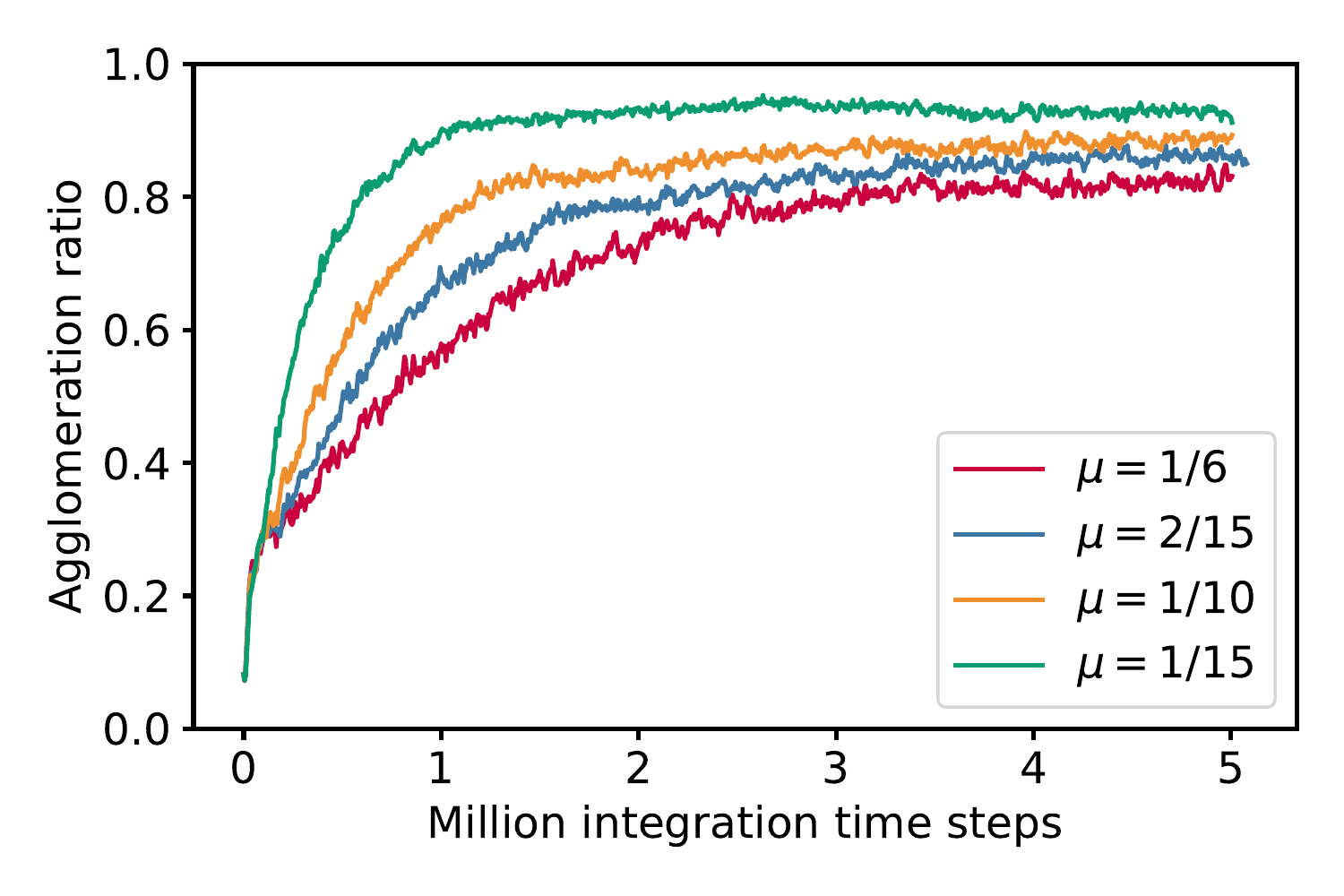}
\caption{The influence of suspending liquid viscosity $\nu$ on the
evolution of particle agglomeration (surface tension coefficient
$\sigma=5\times{}10^{-5}$ and shear rate $\dot\gamma_\mathrm{eff}=6.9
\times 10^{-5}$ ).
  \label{fig:effect_of_viscosity}}
  \end{center}
\end{figure}

Another important factor that influences the suspension's rheology turns out to be the viscosity of the suspending fluid itself. In fact, the momentum transfer from the fluid to the particles is more efficient at high values of the suspending fluid viscosity, and we can expect clusters to be dissolved more easily in high viscosity fluids. One can appreciate this by looking, for example, at the fraction of particles involved in some kind of cluster (agglomeration ratio), as reported in Fig.~\ref{fig:effect_of_viscosity}. The effect of the increasing medium viscosity is seen in the slower relaxation toward the stationary state, and in the higher value of the stationary agglomeration ratio.

\begin{figure}
\begin{center}      
\includegraphics[width=\columnwidth]{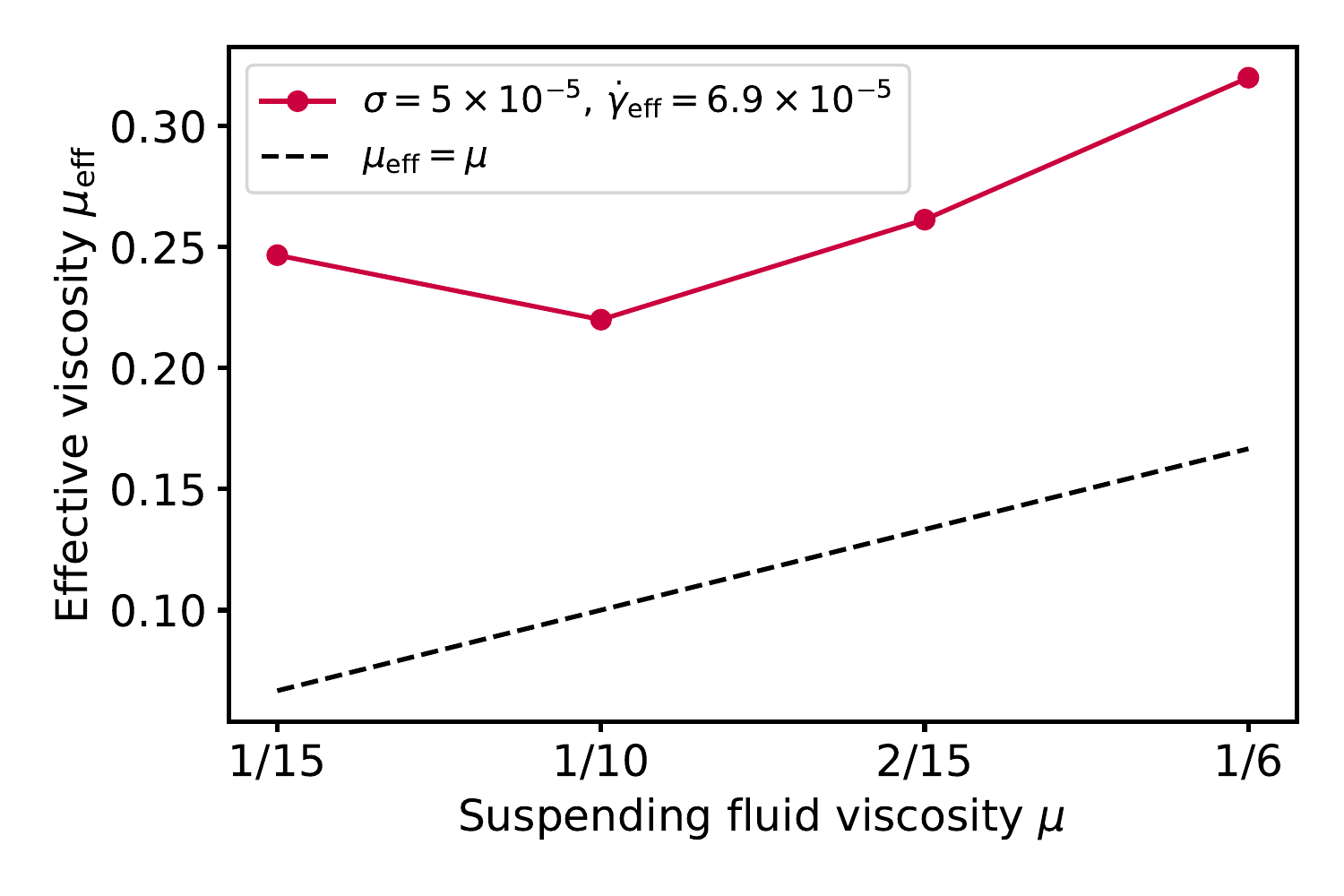}
\end{center}
\caption{The influence of suspending fluid viscosity on the effective 
suspension viscosity, for fixed
$\sigma=5\times{}10^{-5}$ and $\dot\gamma_\mathrm{eff}=6.9 \times
10^{-5}$ ).
}
  \label{fig:absolute_viscosity}
\end{figure}

Therefore, when the viscosity of the suspending medium is high, one can expect 
a decrease in the effective relative viscosity of the suspension due to the 
dissolution of larger clusters. These two, however, are competing effects, and 
it is interesting to look at the cumulative influence on the absolute value of the effective viscosity, as reported in Fig.~\ref{fig:absolute_viscosity}.
The absolute suspension viscosity shows an initial decrease because of the looser clustering, but these are not dissolved efficiently enough at higher suspending fluid viscosities, to keep decreasing the absolute effective viscosity of the suspension. As a result, the effective viscosity of the suspension presents a minimum value, which is located, for this particular value of shear rate and capillary interaction strength, at around $\mu=1/10$. This result shows that, contrarily to what is assumed in empirical models~\cite{einstein1911berichtigung, batchelor1972determination}, the effective viscosity (relative or absolute) does in fact depend on the suspending medium viscosity, as initially also suggested by Konijn and coworkers~\cite{konijn2014experimental}.

\section{Conclusion}
We performed coupled lattice Boltzmann and discrete element
method simulations to investigate the rheology of suspensions of rigid particles in presence of capillary interactions, as a model of particle agglomeration due to molten ashes in fludized bed reactors.
We investigated the influence of shear rate, surface
tension and suspending liquid viscosity on the effective viscosity of the suspension and on the cluster formation.

We observed the characteristic shear-thinning behavior, linked to the presence of large clusters. We found that the morphology of the clusters changes dramatically when the shear rate is increased above a certain threshold, leading to the transformation of the long filaments present at lower shear rates into globular clusters at higher shear rates. The overall behavior at any  given capillary interaction strength is always that of a decrease in the relative viscosity of the suspension, which correlates with the dissolution of the clusters. On the contrary, at fixed shear rates, the suspension shows a non-trivial behavior that includes both a decrease (at low shear rates) and  an increase (at high shear rates) of the relative viscosity, upon increase of the capillary interaction strength. This, again correlates very well with the corresponding decrease (at low shear rates) and increase (at high shear rates) of large clusters. Last, we showed that the 
viscosity of the suspension does also depend on the suspending medium viscosity, indicating a shortcoming of present empirical models.

In this work we were able to provide some connections between the microscopic 
structure and the rheological properties of fluidized suspensions in the presence 
of capillary interactions. While these results represent a step toward a deeper 
understanding of the interaction mechanisms of capillary‐bridge agglomerates, there is still room for improvement in the modelling of these 
systems. Specifically, it would be interesting to include a suitable liquid 
bridge filling model, which describes the portion of liquid involved in the 
bridge~\cite{wu2018effect}. Besides, 
further developments would include the extension of this lattice 
Boltzmann-based approach to turbulent flow problems, given their relevance for 
industrial processes. In this context, a comparison of the simulations 
with experiments is a helpful next step to validate the computational approach. Further work will, for example, include the processing of SEM images of samples at different times in agglomeration experiments to better understand the processes of agglomeration itself. Also, a variation of relevant control parameters  (for example, ash amount, fluidization, additives, \ldots) in the experiments will provide useful information to further improve the simulation of the agglomeration mechanism.

\section*{Appendix}

The definitions of the coefficients $f_1, f_2, f_3$ and $f_4$ for
capillary force model are
\begin{eqnarray}
f_1  = & -0.44507 + 0.050832 \,\theta - 1.1466\,\theta^2 \nonumber\\
     & - (0.1119 + 0.000411\,\theta + 0.1490\,\theta^2 ) \ln \Bar{V}\nonumber\\
     & -(0.012101 + 0.0036456\,\theta + 0.01255\,\theta^2 )(\ln \Bar{V})^2\nonumber\\
     & - (0.0005 + 0.0003505\,\theta + 0.00029076\,\theta^2 )(\ln \Bar{V})^3 \nonumber
\end{eqnarray}

\begin{eqnarray}
f_2  = &  1.9222 - 0.57473\,\theta- 1.2918\,\theta^2\nonumber\\
     & - (0.0668 + 0.1201\,\theta + 0.22574\,\theta^2 )\ln \Bar{V}\nonumber\\
     & - (0.0013375 + 0.0068988\,\theta + 0.01137\,\theta^2 )(\ln \Bar{V})^2\nonumber
\end{eqnarray}

\begin{eqnarray}
f_3  =  &  1.268 - 0.01396\,\theta - 0.23566\,\theta^2\nonumber\\
   & +(0.198 + 0.092\,\theta - 0.06418\,\theta^2 )\ln \Bar{V} \nonumber\\
   & +(0.02232 + 0.02238\,\theta - 0.009853\,\theta^2 )(\ln \Bar{V})^2\nonumber \\
   & +(0.0008585 + 0.001318\,\theta- 0.00053\,\theta^2 )(\ln \Bar{V})^3\nonumber
\end{eqnarray}

\begin{eqnarray}
   f_4  = & -0.010703 + 0.073776\,\theta- 0.34742\,\theta^2 \nonumber\\
    & +(0.03345 + 0.04543\,\theta - 0.09056\,\theta^2 )\ln \Bar{V} \nonumber\\
    &+(0.0018574 + 0.004456\,\theta - 0.006257\,\theta^2 )(\ln \Bar{V})^2,\nonumber
\end{eqnarray}
where $\bar{V}=(4/3) \pi V/ V_p$
\comment{ 

}

\section*{Acknowledgement}

The authors thank Dr. Othmane Aouane for fruitful discussions. Financial support by the German Research Foundation (DFG) within the project HA\,4382/7-1 as well as the computing
time granted by the J\"ulich Supercomputing
Centre (JSC) are highly acknowledged.

\bibliography{references}

\begin{thebibliography}{39}%
\makeatletter
\providecommand \@ifxundefined [1]{%
 \@ifx{#1\undefined}
}%
\providecommand \@ifnum [1]{%
 \ifnum #1\expandafter \@firstoftwo
 \else \expandafter \@secondoftwo
 \fi
}%
\providecommand \@ifx [1]{%
 \ifx #1\expandafter \@firstoftwo
 \else \expandafter \@secondoftwo
 \fi
}%
\providecommand \natexlab [1]{#1}%
\providecommand \enquote  [1]{``#1''}%
\providecommand \bibnamefont  [1]{#1}%
\providecommand \bibfnamefont [1]{#1}%
\providecommand \citenamefont [1]{#1}%
\providecommand \href@noop [0]{\@secondoftwo}%
\providecommand \href [0]{\begingroup \@sanitize@url \@href}%
\providecommand \@href[1]{\@@startlink{#1}\@@href}%
\providecommand \@@href[1]{\endgroup#1\@@endlink}%
\providecommand \@sanitize@url [0]{\catcode `\\12\catcode `\$12\catcode
  `\&12\catcode `\#12\catcode `\^12\catcode `\_12\catcode `\%12\relax}%
\providecommand \@@startlink[1]{}%
\providecommand \@@endlink[0]{}%
\providecommand \url  [0]{\begingroup\@sanitize@url \@url }%
\providecommand \@url [1]{\endgroup\@href {#1}{\urlprefix }}%
\providecommand \urlprefix  [0]{URL }%
\providecommand \Eprint [0]{\href }%
\providecommand \doibase [0]{http://dx.doi.org/}%
\providecommand \selectlanguage [0]{\@gobble}%
\providecommand \bibinfo  [0]{\@secondoftwo}%
\providecommand \bibfield  [0]{\@secondoftwo}%
\providecommand \translation [1]{[#1]}%
\providecommand \BibitemOpen [0]{}%
\providecommand \bibitemStop [0]{}%
\providecommand \bibitemNoStop [0]{.\EOS\space}%
\providecommand \EOS [0]{\spacefactor3000\relax}%
\providecommand \BibitemShut  [1]{\csname bibitem#1\endcsname}%
\let\auto@bib@innerbib\@empty
\bibitem [{\citenamefont {Visser}\ \emph {et~al.}(2004)\citenamefont {Visser},
  \citenamefont {Kiel},\ and\ \citenamefont {Veringa}}]{visser2004influence}%
  \BibitemOpen
  \bibfield  {author} {\bibinfo {author} {\bibfnamefont {H.~J.~M.}\
  \bibnamefont {Visser}}, \bibinfo {author} {\bibfnamefont {J.}~\bibnamefont
  {Kiel}}, \ and\ \bibinfo {author} {\bibfnamefont {H.}~\bibnamefont
  {Veringa}},\ }\href@noop {} {\emph {\bibinfo {title} {The influence of fuel
  composition on agglomeration behaviour in fluidised-bed combustion}}}\
  (\bibinfo  {publisher} {Energy research Centre of the Netherlands ECN
  Delft},\ \bibinfo {year} {2004})\BibitemShut {NoStop}%
\bibitem [{\citenamefont {Gatternig}\ and\ \citenamefont
  {Karl}(2015)}]{gatternig2015investigations}%
  \BibitemOpen
  \bibfield  {author} {\bibinfo {author} {\bibfnamefont {B.}~\bibnamefont
  {Gatternig}}\ and\ \bibinfo {author} {\bibfnamefont {J.}~\bibnamefont
  {Karl}},\ }\href {\doibase 10.1021/ef502658b} {\bibfield  {journal} {\bibinfo
   {journal} {Energy Fuels}\ }\textbf {\bibinfo {volume} {29}},\ \bibinfo
  {pages} {931} (\bibinfo {year} {2015})}\BibitemShut {NoStop}%
\bibitem [{\citenamefont {Khan}\ \emph {et~al.}(2009)\citenamefont {Khan},
  \citenamefont {de~Jong}, \citenamefont {Jansens},\ and\ \citenamefont
  {Spliethoff}}]{khan2009biomass}%
  \BibitemOpen
  \bibfield  {author} {\bibinfo {author} {\bibfnamefont {A.}~\bibnamefont
  {Khan}}, \bibinfo {author} {\bibfnamefont {W.}~\bibnamefont {de~Jong}},
  \bibinfo {author} {\bibfnamefont {P.}~\bibnamefont {Jansens}}, \ and\
  \bibinfo {author} {\bibfnamefont {H.}~\bibnamefont {Spliethoff}},\ }\href
  {\doibase 10.1016/j.fuproc.2008.07.012} {\bibfield  {journal} {\bibinfo
  {journal} {Fuel Process. Technol.}\ }\textbf {\bibinfo {volume} {90}},\
  \bibinfo {pages} {21} (\bibinfo {year} {2009})}\BibitemShut {NoStop}%
\bibitem [{\citenamefont {Ergudenler}\ and\ \citenamefont
  {Ghaly}(1993)}]{ergudenler1993agglomeration}%
  \BibitemOpen
  \bibfield  {author} {\bibinfo {author} {\bibfnamefont {A.}~\bibnamefont
  {Ergudenler}}\ and\ \bibinfo {author} {\bibfnamefont {A.}~\bibnamefont
  {Ghaly}},\ }\href {\doibase 10.1016/0961-9534(93)90034-2} {\bibfield
  {journal} {\bibinfo  {journal} {Biomass Bioenergy}\ }\textbf {\bibinfo
  {volume} {4}},\ \bibinfo {pages} {135} (\bibinfo {year} {1993})}\BibitemShut
  {NoStop}%
\bibitem [{\citenamefont {{\"O}hman}\ \emph {et~al.}(2000)\citenamefont
  {{\"O}hman}, \citenamefont {Nordin}, \citenamefont {Skrifvars}, \citenamefont
  {Backman},\ and\ \citenamefont {Hupa}}]{ohman2000bed}%
  \BibitemOpen
  \bibfield  {author} {\bibinfo {author} {\bibfnamefont {M.}~\bibnamefont
  {{\"O}hman}}, \bibinfo {author} {\bibfnamefont {A.}~\bibnamefont {Nordin}},
  \bibinfo {author} {\bibfnamefont {B.-J.}\ \bibnamefont {Skrifvars}}, \bibinfo
  {author} {\bibfnamefont {R.}~\bibnamefont {Backman}}, \ and\ \bibinfo
  {author} {\bibfnamefont {M.}~\bibnamefont {Hupa}},\ }\href {\doibase
  10.1021/ef990107b} {\bibfield  {journal} {\bibinfo  {journal} {Energy Fuels}\
  }\textbf {\bibinfo {volume} {14}},\ \bibinfo {pages} {169} (\bibinfo {year}
  {2000})}\BibitemShut {NoStop}%
\bibitem [{\citenamefont {Olofsson}\ \emph {et~al.}(2002)\citenamefont
  {Olofsson}, \citenamefont {Ye}, \citenamefont {Bjerle},\ and\ \citenamefont
  {Andersson}}]{olofsson2002bed}%
  \BibitemOpen
  \bibfield  {author} {\bibinfo {author} {\bibfnamefont {G.}~\bibnamefont
  {Olofsson}}, \bibinfo {author} {\bibfnamefont {Z.}~\bibnamefont {Ye}},
  \bibinfo {author} {\bibfnamefont {I.}~\bibnamefont {Bjerle}}, \ and\ \bibinfo
  {author} {\bibfnamefont {A.}~\bibnamefont {Andersson}},\ }\href {\doibase
  10.1021/ie010274a} {\bibfield  {journal} {\bibinfo  {journal} {Ind. Eng.
  Chem. Res.}\ }\textbf {\bibinfo {volume} {41}},\ \bibinfo {pages} {2888}
  (\bibinfo {year} {2002})}\BibitemShut {NoStop}%
\bibitem [{\citenamefont {Koos}\ and\ \citenamefont
  {Willenbacher}(2011)}]{koos2011capillary}%
  \BibitemOpen
  \bibfield  {author} {\bibinfo {author} {\bibfnamefont {E.}~\bibnamefont
  {Koos}}\ and\ \bibinfo {author} {\bibfnamefont {N.}~\bibnamefont
  {Willenbacher}},\ }\href {\doibase 10.1126/science.1199243} {\bibfield
  {journal} {\bibinfo  {journal} {Science}\ }\textbf {\bibinfo {volume}
  {331}},\ \bibinfo {pages} {897} (\bibinfo {year} {2011})}\BibitemShut
  {NoStop}%
\bibitem [{\citenamefont {Koos}\ \emph {et~al.}(2012)\citenamefont {Koos},
  \citenamefont {Johannsmeier}, \citenamefont {Schwebler},\ and\ \citenamefont
  {Willenbacher}}]{koos2012tuning}%
  \BibitemOpen
  \bibfield  {author} {\bibinfo {author} {\bibfnamefont {E.}~\bibnamefont
  {Koos}}, \bibinfo {author} {\bibfnamefont {J.}~\bibnamefont {Johannsmeier}},
  \bibinfo {author} {\bibfnamefont {L.}~\bibnamefont {Schwebler}}, \ and\
  \bibinfo {author} {\bibfnamefont {N.}~\bibnamefont {Willenbacher}},\ }\href
  {\doibase 10.1039/C2SM25681A} {\bibfield  {journal} {\bibinfo  {journal}
  {Soft Matter}\ }\textbf {\bibinfo {volume} {8}},\ \bibinfo {pages} {6620}
  (\bibinfo {year} {2012})}\BibitemShut {NoStop}%
\bibitem [{\citenamefont {Herminghaus}(2005)}]{herminghaus2005}%
  \BibitemOpen
  \bibfield  {author} {\bibinfo {author} {\bibfnamefont {S.}~\bibnamefont
  {Herminghaus}},\ }\href {\doibase 10.1080/00018730500167855} {\bibfield
  {journal} {\bibinfo  {journal} {Adv. Phys.}\ }\textbf {\bibinfo {volume}
  {54}},\ \bibinfo {pages} {221} (\bibinfo {year} {2005})}\BibitemShut
  {NoStop}%
\bibitem [{\citenamefont {Cho}\ \emph {et~al.}(2006)\citenamefont {Cho},
  \citenamefont {Mattisson},\ and\ \citenamefont
  {Lyngfelt}}]{cho2006defluidization}%
  \BibitemOpen
  \bibfield  {author} {\bibinfo {author} {\bibfnamefont {P.}~\bibnamefont
  {Cho}}, \bibinfo {author} {\bibfnamefont {T.}~\bibnamefont {Mattisson}}, \
  and\ \bibinfo {author} {\bibfnamefont {A.}~\bibnamefont {Lyngfelt}},\ }\href
  {\doibase 10.1021/ie050484d} {\bibfield  {journal} {\bibinfo  {journal} {Ind.
  Eng. Chem. Res.}\ }\textbf {\bibinfo {volume} {45}},\ \bibinfo {pages} {968}
  (\bibinfo {year} {2006})}\BibitemShut {NoStop}%
\bibitem [{\citenamefont {Fryda}\ \emph {et~al.}(2008)\citenamefont {Fryda},
  \citenamefont {Panopoulos},\ and\ \citenamefont
  {Kakaras}}]{fryda2008agglomeration}%
  \BibitemOpen
  \bibfield  {author} {\bibinfo {author} {\bibfnamefont {L.}~\bibnamefont
  {Fryda}}, \bibinfo {author} {\bibfnamefont {K.~D.}\ \bibnamefont
  {Panopoulos}}, \ and\ \bibinfo {author} {\bibfnamefont {E.}~\bibnamefont
  {Kakaras}},\ }\href {\doibase 10.1016/j.powtec.2007.05.022} {\bibfield
  {journal} {\bibinfo  {journal} {Powder Technol.}\ }\textbf {\bibinfo {volume}
  {181}},\ \bibinfo {pages} {307} (\bibinfo {year} {2008})}\BibitemShut
  {NoStop}%
\bibitem [{\citenamefont {Scala}\ \emph {et~al.}(2006)\citenamefont {Scala},
  \citenamefont {Chirone},\ and\ \citenamefont
  {Salatino}}]{scala2006combustion}%
  \BibitemOpen
  \bibfield  {author} {\bibinfo {author} {\bibfnamefont {F.}~\bibnamefont
  {Scala}}, \bibinfo {author} {\bibfnamefont {R.}~\bibnamefont {Chirone}}, \
  and\ \bibinfo {author} {\bibfnamefont {P.}~\bibnamefont {Salatino}},\ }\href
  {\doibase 10.1021/ef050102g} {\bibfield  {journal} {\bibinfo  {journal}
  {Energy Fuels}\ }\textbf {\bibinfo {volume} {20}},\ \bibinfo {pages} {91}
  (\bibinfo {year} {2006})}\BibitemShut {NoStop}%
\bibitem [{\citenamefont {Sch{\"u}gerl}(1971)}]{schugerl1971rheological}%
  \BibitemOpen
  \bibfield  {author} {\bibinfo {author} {\bibfnamefont {K.}~\bibnamefont
  {Sch{\"u}gerl}},\ }\href@noop {} {\bibfield  {journal} {\bibinfo  {journal}
  {Fluidization}\ ,\ \bibinfo {pages} {261}} (\bibinfo {year}
  {1971})}\BibitemShut {NoStop}%
\bibitem [{\citenamefont {Van~Kao}\ \emph {et~al.}(1975)\citenamefont
  {Van~Kao}, \citenamefont {Nielsen},\ and\ \citenamefont
  {Hill}}]{van1975rheology}%
  \BibitemOpen
  \bibfield  {author} {\bibinfo {author} {\bibfnamefont {S.}~\bibnamefont
  {Van~Kao}}, \bibinfo {author} {\bibfnamefont {L.~E.}\ \bibnamefont
  {Nielsen}}, \ and\ \bibinfo {author} {\bibfnamefont {C.~T.}\ \bibnamefont
  {Hill}},\ }\href {\doibase 10.1016/0021-9797(75)90052-1} {\bibfield
  {journal} {\bibinfo  {journal} {J. Colloid Interface Sci.}\ }\textbf
  {\bibinfo {volume} {53}},\ \bibinfo {pages} {367} (\bibinfo {year}
  {1975})}\BibitemShut {NoStop}%
\bibitem [{\citenamefont {McCulfor}\ \emph {et~al.}(2011)\citenamefont
  {McCulfor}, \citenamefont {Himes},\ and\ \citenamefont
  {Anklam}}]{mcculfor2011effects}%
  \BibitemOpen
  \bibfield  {author} {\bibinfo {author} {\bibfnamefont {J.}~\bibnamefont
  {McCulfor}}, \bibinfo {author} {\bibfnamefont {P.}~\bibnamefont {Himes}}, \
  and\ \bibinfo {author} {\bibfnamefont {M.~R.}\ \bibnamefont {Anklam}},\
  }\href {\doibase 10.1002/aic.12451} {\bibfield  {journal} {\bibinfo
  {journal} {AIChE J.}\ }\textbf {\bibinfo {volume} {57}},\ \bibinfo {pages}
  {2334} (\bibinfo {year} {2011})}\BibitemShut {NoStop}%
\bibitem [{\citenamefont {Mikami}\ \emph {et~al.}(1998)\citenamefont {Mikami},
  \citenamefont {Kamiya},\ and\ \citenamefont {Horio}}]{mikami1998numerical}%
  \BibitemOpen
  \bibfield  {author} {\bibinfo {author} {\bibfnamefont {T.}~\bibnamefont
  {Mikami}}, \bibinfo {author} {\bibfnamefont {H.}~\bibnamefont {Kamiya}}, \
  and\ \bibinfo {author} {\bibfnamefont {M.}~\bibnamefont {Horio}},\ }\href
  {\doibase 10.1016/S0009-2509(97)00325-4} {\bibfield  {journal} {\bibinfo
  {journal} {Chem. Eng. Sci.}\ }\textbf {\bibinfo {volume} {53}},\ \bibinfo
  {pages} {1927} (\bibinfo {year} {1998})}\BibitemShut {NoStop}%
\bibitem [{\citenamefont {Wu}\ \emph {et~al.}(2018)\citenamefont {Wu},
  \citenamefont {Khinast},\ and\ \citenamefont {Radl}}]{wu2018effect}%
  \BibitemOpen
  \bibfield  {author} {\bibinfo {author} {\bibfnamefont {M.}~\bibnamefont
  {Wu}}, \bibinfo {author} {\bibfnamefont {J.~G.}\ \bibnamefont {Khinast}}, \
  and\ \bibinfo {author} {\bibfnamefont {S.}~\bibnamefont {Radl}},\ }\href
  {\doibase 10.1002/aic.15947} {\bibfield  {journal} {\bibinfo  {journal}
  {AIChE J.}\ }\textbf {\bibinfo {volume} {64}},\ \bibinfo {pages} {437}
  (\bibinfo {year} {2018})}\BibitemShut {NoStop}%
\bibitem [{\citenamefont {Zhang}\ \emph {et~al.}(2017)\citenamefont {Zhang},
  \citenamefont {Zhao}, \citenamefont {Lu}, \citenamefont {Ge}, \citenamefont
  {Wang},\ and\ \citenamefont {Duan}}]{zhang2017assessment}%
  \BibitemOpen
  \bibfield  {author} {\bibinfo {author} {\bibfnamefont {Y.}~\bibnamefont
  {Zhang}}, \bibinfo {author} {\bibfnamefont {Y.}~\bibnamefont {Zhao}},
  \bibinfo {author} {\bibfnamefont {L.}~\bibnamefont {Lu}}, \bibinfo {author}
  {\bibfnamefont {W.}~\bibnamefont {Ge}}, \bibinfo {author} {\bibfnamefont
  {J.}~\bibnamefont {Wang}}, \ and\ \bibinfo {author} {\bibfnamefont
  {C.}~\bibnamefont {Duan}},\ }\href {\doibase 10.1016/j.ces.2016.11.028}
  {\bibfield  {journal} {\bibinfo  {journal} {Chem. Eng. Sci.}\ }\textbf
  {\bibinfo {volume} {160}},\ \bibinfo {pages} {106} (\bibinfo {year}
  {2017})}\BibitemShut {NoStop}%
\bibitem [{\citenamefont {Roy}\ \emph {et~al.}(2017)\citenamefont {Roy},
  \citenamefont {Luding},\ and\ \citenamefont {Weinhart}}]{roy2017}%
  \BibitemOpen
  \bibfield  {author} {\bibinfo {author} {\bibfnamefont {S.}~\bibnamefont
  {Roy}}, \bibinfo {author} {\bibfnamefont {S.}~\bibnamefont {Luding}}, \ and\
  \bibinfo {author} {\bibfnamefont {T.}~\bibnamefont {Weinhart}},\ }\href
  {\doibase 10.1088/1367-2630/aa6141} {\bibfield  {journal} {\bibinfo
  {journal} {New J. Phys.}\ }\textbf {\bibinfo {volume} {19}},\ \bibinfo
  {pages} {043014} (\bibinfo {year} {2017})}\BibitemShut {NoStop}%
\bibitem [{\citenamefont {Koos}\ \emph {et~al.}(2014)\citenamefont {Koos},
  \citenamefont {Kannowade},\ and\ \citenamefont
  {Willenbacher}}]{koos2014restructuring}%
  \BibitemOpen
  \bibfield  {author} {\bibinfo {author} {\bibfnamefont {E.}~\bibnamefont
  {Koos}}, \bibinfo {author} {\bibfnamefont {W.}~\bibnamefont {Kannowade}}, \
  and\ \bibinfo {author} {\bibfnamefont {N.}~\bibnamefont {Willenbacher}},\
  }\href {\doibase 10.1007/s00397-014-0805-z} {\bibfield  {journal} {\bibinfo
  {journal} {Rheol. Acta}\ }\textbf {\bibinfo {volume} {53}},\ \bibinfo {pages}
  {947} (\bibinfo {year} {2014})}\BibitemShut {NoStop}%
\bibitem [{\citenamefont {Benzi}\ \emph {et~al.}(1992)\citenamefont {Benzi},
  \citenamefont {Succi},\ and\ \citenamefont {Vergassola}}]{benzi1992lattice}%
  \BibitemOpen
  \bibfield  {author} {\bibinfo {author} {\bibfnamefont {R.}~\bibnamefont
  {Benzi}}, \bibinfo {author} {\bibfnamefont {S.}~\bibnamefont {Succi}}, \ and\
  \bibinfo {author} {\bibfnamefont {M.}~\bibnamefont {Vergassola}},\ }\href
  {\doibase 10.1016/0370-1573(92)90090-M} {\bibfield  {journal} {\bibinfo
  {journal} {Phys. Rep.}\ }\textbf {\bibinfo {volume} {222}},\ \bibinfo {pages}
  {145} (\bibinfo {year} {1992})}\BibitemShut {NoStop}%
\bibitem [{\citenamefont {Bhatnagar}\ \emph {et~al.}(1954)\citenamefont
  {Bhatnagar}, \citenamefont {Gross},\ and\ \citenamefont
  {Krook}}]{bhatnagar1954model}%
  \BibitemOpen
  \bibfield  {author} {\bibinfo {author} {\bibfnamefont {P.~L.}\ \bibnamefont
  {Bhatnagar}}, \bibinfo {author} {\bibfnamefont {E.~P.}\ \bibnamefont
  {Gross}}, \ and\ \bibinfo {author} {\bibfnamefont {M.}~\bibnamefont
  {Krook}},\ }\href {\doibase 10.1103/PhysRev.94.511} {\bibfield  {journal}
  {\bibinfo  {journal} {Phys. Rev.}\ }\textbf {\bibinfo {volume} {94}},\
  \bibinfo {pages} {511} (\bibinfo {year} {1954})}\BibitemShut {NoStop}%
\bibitem [{\citenamefont {Ladd}(1994)}]{ladd1994numerical}%
  \BibitemOpen
  \bibfield  {author} {\bibinfo {author} {\bibfnamefont {A.~J.~C.}\
  \bibnamefont {Ladd}},\ }\href {\doibase 10.1017/S0022112094001771} {\bibfield
   {journal} {\bibinfo  {journal} {J. Fluid Mech.}\ }\textbf {\bibinfo {volume}
  {271}},\ \bibinfo {pages} {285} (\bibinfo {year} {1994})}\BibitemShut
  {NoStop}%
\bibitem [{\citenamefont {Aidun}\ \emph {et~al.}(1998)\citenamefont {Aidun},
  \citenamefont {Lu},\ and\ \citenamefont {Ding}}]{aidun1998direct}%
  \BibitemOpen
  \bibfield  {author} {\bibinfo {author} {\bibfnamefont {C.~K.}\ \bibnamefont
  {Aidun}}, \bibinfo {author} {\bibfnamefont {Y.}~\bibnamefont {Lu}}, \ and\
  \bibinfo {author} {\bibfnamefont {E.}~\bibnamefont {Ding}},\ }\href {\doibase
  10.1017/S0022112098002493} {\bibfield  {journal} {\bibinfo  {journal} {J.
  Fluid Mech.}\ }\textbf {\bibinfo {volume} {373}},\ \bibinfo {pages} {287}
  (\bibinfo {year} {1998})}\BibitemShut {NoStop}%
\bibitem [{\citenamefont {Ladd}\ and\ \citenamefont
  {Verberg}(2001)}]{ladd2001lattice}%
  \BibitemOpen
  \bibfield  {author} {\bibinfo {author} {\bibfnamefont {A.}~\bibnamefont
  {Ladd}}\ and\ \bibinfo {author} {\bibfnamefont {R.}~\bibnamefont {Verberg}},\
  }\href {\doibase 10.1023/A:1010414013942} {\bibfield  {journal} {\bibinfo
  {journal} {J. Stat. Phys.}\ }\textbf {\bibinfo {volume} {104}},\ \bibinfo
  {pages} {1191} (\bibinfo {year} {2001})}\BibitemShut {NoStop}%
\bibitem [{\citenamefont {Hertz}(1882)}]{hertz1882ueber}%
  \BibitemOpen
  \bibfield  {author} {\bibinfo {author} {\bibfnamefont {H.}~\bibnamefont
  {Hertz}},\ }\href {\doibase 10.1515/9783112342404-004} {\bibfield  {journal}
  {\bibinfo  {journal} {J. Reine Angew. Math.}\ }\textbf {\bibinfo {volume}
  {92}},\ \bibinfo {pages} {156} (\bibinfo {year} {1882})}\BibitemShut
  {NoStop}%
\bibitem [{\citenamefont {Willett}\ \emph {et~al.}(2000)\citenamefont
  {Willett}, \citenamefont {Adams}, \citenamefont {Johnson},\ and\
  \citenamefont {Seville}}]{willett2000capillary}%
  \BibitemOpen
  \bibfield  {author} {\bibinfo {author} {\bibfnamefont {C.~D.}\ \bibnamefont
  {Willett}}, \bibinfo {author} {\bibfnamefont {M.~J.}\ \bibnamefont {Adams}},
  \bibinfo {author} {\bibfnamefont {S.~A.}\ \bibnamefont {Johnson}}, \ and\
  \bibinfo {author} {\bibfnamefont {J.~P.}\ \bibnamefont {Seville}},\ }\href
  {\doibase 10.1021/la000657y} {\bibfield  {journal} {\bibinfo  {journal}
  {Langmuir}\ }\textbf {\bibinfo {volume} {16}},\ \bibinfo {pages} {9396}
  (\bibinfo {year} {2000})}\BibitemShut {NoStop}%
\bibitem [{\citenamefont {Yang}\ \emph {et~al.}(2021)\citenamefont {Yang},
  \citenamefont {Sega},\ and\ \citenamefont {Harting}}]{yang2020capillary}%
  \BibitemOpen
  \bibfield  {author} {\bibinfo {author} {\bibfnamefont {L.}~\bibnamefont
  {Yang}}, \bibinfo {author} {\bibfnamefont {M.}~\bibnamefont {Sega}}, \ and\
  \bibinfo {author} {\bibfnamefont {J.}~\bibnamefont {Harting}},\ }\href
  {\doibase https://doi.org/10.1002/aic.17350} {\bibfield  {journal} {\bibinfo
  {journal} {AIChE J.}\ }\textbf {\bibinfo {volume} {67}},\ \bibinfo {pages}
  {e17350} (\bibinfo {year} {2021})}\BibitemShut {NoStop}%
\bibitem [{\citenamefont {Lees}\ and\ \citenamefont
  {Edwards}(1972)}]{lees1972computer}%
  \BibitemOpen
  \bibfield  {author} {\bibinfo {author} {\bibfnamefont {A.}~\bibnamefont
  {Lees}}\ and\ \bibinfo {author} {\bibfnamefont {S.}~\bibnamefont {Edwards}},\
  }\href {\doibase 10.1088/0022-3719/5/15/006} {\bibfield  {journal} {\bibinfo
  {journal} {J. Phys. C: Solid State Phys.}\ }\textbf {\bibinfo {volume} {5}},\
  \bibinfo {pages} {1921} (\bibinfo {year} {1972})}\BibitemShut {NoStop}%
\bibitem [{\citenamefont {Wagner}\ and\ \citenamefont
  {Pagonabarraga}(2002)}]{wagner2002lees}%
  \BibitemOpen
  \bibfield  {author} {\bibinfo {author} {\bibfnamefont {A.~J.}\ \bibnamefont
  {Wagner}}\ and\ \bibinfo {author} {\bibfnamefont {I.}~\bibnamefont
  {Pagonabarraga}},\ }\href {\doibase 10.1023/A:1014595628808} {\bibfield
  {journal} {\bibinfo  {journal} {J. Stat. Phys.}\ }\textbf {\bibinfo {volume}
  {107}},\ \bibinfo {pages} {521} (\bibinfo {year} {2002})}\BibitemShut
  {NoStop}%
\bibitem [{\citenamefont {Harting}\ \emph {et~al.}(2004)\citenamefont
  {Harting}, \citenamefont {Venturoli},\ and\ \citenamefont {Coveney}}]{HVC04}%
  \BibitemOpen
  \bibfield  {author} {\bibinfo {author} {\bibfnamefont {J.}~\bibnamefont
  {Harting}}, \bibinfo {author} {\bibfnamefont {M.}~\bibnamefont {Venturoli}},
  \ and\ \bibinfo {author} {\bibfnamefont {P.~V.}\ \bibnamefont {Coveney}},\
  }\href {\doibase 10.1098/rsta.2004.1402} {\bibfield  {journal} {\bibinfo
  {journal} {Phil. Trans. R. Soc. London Series A}\ }\textbf {\bibinfo {volume}
  {362}},\ \bibinfo {pages} {1703} (\bibinfo {year} {2004})}\BibitemShut
  {NoStop}%
\bibitem [{\citenamefont {Lallemand}\ and\ \citenamefont
  {Luo}(2000)}]{lallemand2000theory}%
  \BibitemOpen
  \bibfield  {author} {\bibinfo {author} {\bibfnamefont {P.}~\bibnamefont
  {Lallemand}}\ and\ \bibinfo {author} {\bibfnamefont {L.-S.}\ \bibnamefont
  {Luo}},\ }\href {\doibase 10.1103/PhysRevE.61.6546} {\bibfield  {journal}
  {\bibinfo  {journal} {Phys. Rev. E}\ }\textbf {\bibinfo {volume} {61}},\
  \bibinfo {pages} {6546} (\bibinfo {year} {2000})}\BibitemShut {NoStop}%
\bibitem [{\citenamefont {Huang}\ \emph {et~al.}(2012)\citenamefont {Huang},
  \citenamefont {Yang}, \citenamefont {Krafczyk},\ and\ \citenamefont
  {Lu}}]{huang2012rotation}%
  \BibitemOpen
  \bibfield  {author} {\bibinfo {author} {\bibfnamefont {H.}~\bibnamefont
  {Huang}}, \bibinfo {author} {\bibfnamefont {X.}~\bibnamefont {Yang}},
  \bibinfo {author} {\bibfnamefont {M.}~\bibnamefont {Krafczyk}}, \ and\
  \bibinfo {author} {\bibfnamefont {X.-Y.}\ \bibnamefont {Lu}},\ }\href
  {\doibase 10.1017/jfm.2011.519} {\bibfield  {journal} {\bibinfo  {journal}
  {J. Fluid Mech.}\ }\textbf {\bibinfo {volume} {692}},\ \bibinfo {pages} {369}
  (\bibinfo {year} {2012})}\BibitemShut {NoStop}%
\bibitem [{\citenamefont {Janoschek}(2013)}]{janoschek2013mesoscopic}%
  \BibitemOpen
  \bibfield  {author} {\bibinfo {author} {\bibfnamefont {F.}~\bibnamefont
  {Janoschek}},\ }\emph {\bibinfo {title} {Mesoscopic simulation of blood and
  general suspensions in flow}},\ \href {\doibase 10.6100/IR761379} {Ph.D.
  thesis},\ \bibinfo  {school} {Technische Universiteit Eindhoven}, \bibinfo
  {address} {Eindhoven, The Netherlands} (\bibinfo {year} {2013})\BibitemShut
  {NoStop}%
\bibitem [{\citenamefont {Sega}\ \emph {et~al.}(2018)\citenamefont {Sega},
  \citenamefont {Hantal}, \citenamefont {F{\'a}bi{\'a}n},\ and\ \citenamefont
  {Jedlovszky}}]{sega2018pytim}%
  \BibitemOpen
  \bibfield  {author} {\bibinfo {author} {\bibfnamefont {M.}~\bibnamefont
  {Sega}}, \bibinfo {author} {\bibfnamefont {G.}~\bibnamefont {Hantal}},
  \bibinfo {author} {\bibfnamefont {B.}~\bibnamefont {F{\'a}bi{\'a}n}}, \ and\
  \bibinfo {author} {\bibfnamefont {P.}~\bibnamefont {Jedlovszky}},\ }\href
  {\doibase 10.1002/jcc.25384} {\enquote {\bibinfo {title} {Pytim: A python
  package for the interfacial analysis of molecular simulations},}\ } (\bibinfo
  {year} {2018})\BibitemShut {NoStop}%
\bibitem [{\citenamefont {Hoffmann}\ \emph {et~al.}(2014)\citenamefont
  {Hoffmann}, \citenamefont {Koos},\ and\ \citenamefont
  {Willenbacher}}]{hoffmann2014using}%
  \BibitemOpen
  \bibfield  {author} {\bibinfo {author} {\bibfnamefont {S.}~\bibnamefont
  {Hoffmann}}, \bibinfo {author} {\bibfnamefont {E.}~\bibnamefont {Koos}}, \
  and\ \bibinfo {author} {\bibfnamefont {N.}~\bibnamefont {Willenbacher}},\
  }\href {\doibase 10.1016/j.foodhyd.2014.01.027} {\bibfield  {journal}
  {\bibinfo  {journal} {Food Hydrocolloids}\ }\textbf {\bibinfo {volume}
  {40}},\ \bibinfo {pages} {44} (\bibinfo {year} {2014})}\BibitemShut {NoStop}%
\bibitem [{\citenamefont {Einstein}(1911)}]{einstein1911berichtigung}%
  \BibitemOpen
  \bibfield  {author} {\bibinfo {author} {\bibfnamefont {A.}~\bibnamefont
  {Einstein}},\ }\href {\doibase 10.1002/andp.19113390313} {\bibfield
  {journal} {\bibinfo  {journal} {Ann. Phys.}\ }\textbf {\bibinfo {volume}
  {339}},\ \bibinfo {pages} {591} (\bibinfo {year} {1911})}\BibitemShut
  {NoStop}%
\bibitem [{\citenamefont {Batchelor}\ and\ \citenamefont
  {Green}(1972)}]{batchelor1972determination}%
  \BibitemOpen
  \bibfield  {author} {\bibinfo {author} {\bibfnamefont {G.}~\bibnamefont
  {Batchelor}}\ and\ \bibinfo {author} {\bibfnamefont {J.}~\bibnamefont
  {Green}},\ }\href {\doibase 10.1017/S0022112072002435} {\bibfield  {journal}
  {\bibinfo  {journal} {J. Fluid Mech.}\ }\textbf {\bibinfo {volume} {56}},\
  \bibinfo {pages} {401} (\bibinfo {year} {1972})}\BibitemShut {NoStop}%
\bibitem [{\citenamefont {Konijn}\ \emph {et~al.}(2014)\citenamefont {Konijn},
  \citenamefont {Sanderink},\ and\ \citenamefont
  {Kruyt}}]{konijn2014experimental}%
  \BibitemOpen
  \bibfield  {author} {\bibinfo {author} {\bibfnamefont {B.}~\bibnamefont
  {Konijn}}, \bibinfo {author} {\bibfnamefont {O.}~\bibnamefont {Sanderink}}, \
  and\ \bibinfo {author} {\bibfnamefont {N.~P.}\ \bibnamefont {Kruyt}},\ }\href
  {\doibase 10.1016/j.powtec.2014.05.044} {\bibfield  {journal} {\bibinfo
  {journal} {Powder Technol.}\ }\textbf {\bibinfo {volume} {266}},\ \bibinfo
  {pages} {61} (\bibinfo {year} {2014})}\BibitemShut {NoStop}%
\end{thebibliography}%

\clearpage
\onecolumngrid

\end{document}